# Magnetization and Josephson current coupling drives the Josephson diode effect in $FeTe_{0.55}Se_{0.45}$ van der Waals junctions

Seung-Hyun Shin[1,†], Cheong-Eung Ahn[1,†], Chang-Won Choi[2, 6], Taekoo Oh[3], Si-Young Choi[2, 6, 7], Genda Gu[4], Gil Young Cho[5, 8,*] and Gil-Ho Lee[1,*]

[1] Department of Physics, POSTECH, Pohang, South Korea
[2] Department of Materials Science and Engineering, POSTECH, Pohang, South Korea
[3] Department of Physics, Soongsil University, Seoul, South Korea
[4] Condensed Matter Physics and Materials Science Department, Brookhaven National Laboratory, Upton, NY, USA
[5] Department of Physics, Korea Advanced Institute of Science and Technology, Daejeon, South Korea
[6] Center for Van der Waals Quantum Solids, Institute for Basic Science (IBS), Pohang, South Korea
[7] Department of Semiconductor Engineering, POSTECH, Pohang, South Korea
[8] Center for Artificial Low Dimensional Electronic Systems, Institute for Basic Science, Pohang, South Korea

†*These authors equally contributed.*

*Correspondence to Gil Young Cho (gilyoungcho@kaist.ac.kr) & Gil-Ho Lee (lghman@postech.ac.kr)

ABSTRACT: The Josephson diode effect (JDE) reflects symmetry breaking in the current-phase relation (CPR) of Josephson junctions, yet establishing its microscopic origin is often challenging. We investigate the mechanism of the JDE in van der Waals Josephson junctions based on $FeTe_{0.55}Se_{0.45}$, a platform where superconductivity, topology, and magnetism may coexist at the interface. Our devices exhibit a robust zero-field diode response with asymmetric switching currents. Magnetic-field measurements reveal a hysteretic Fraunhofer pattern and magnetic-resonance-induced voltage feature, evidencing interfacial magnetism coupled to the Josephson current. Microwave response further shows half-integer Shapiro steps with asymmetric evolution, demonstrating contributions from both a second harmonic and a magnetization-induced phase shift in CPR. Numerical simulations combining magnetization dynamics and Josephson phase dynamics reproduce these features, confirming that the JDE originates from interfacial coupling between magnetization and Josephson current. This work provides a transport-based route to pinpoint the microscopic mechanism of the JDE, establishing a platform for exploring magnetically active interfacial superconductivity, and enables the design of nonreciprocal superconducting devices in clean, interface-engineered van der Waals systems.

Nonreciprocal charge transport, in which the response depends on current direction, is a hallmark of symmetry breaking and underpins the operation of diodes[1, 2]. In superconducting systems, such nonreciprocity can emerge without dissipation, a phenomenon broadly referred to as the superconducting diode effect (SDE)[3, 4]. When this behavior originates from an asymmetry in the current-phase relation (CPR) of a Josephson junction, it is termed the Josephson diode effect (JDE)[5-9], reflecting broken inversion and time-reversal symmetries at the junction. Although the JDE has been reported in a variety of systems, its microscopic origin remains difficult to disentangle, as multiple interfacial mechanisms such as spin-orbit coupling (SOC)[8, 10, 11], magnetism[12, 13], and higher-order tunneling processes[14] can coexist and contribute simultaneously. Identifying the dominant mechanism and clarifying how it modifies the CPR, therefore, remains central challenges.

$FeTe_{0.55}Se_{0.45}$ (FTS) provides a suitable platform to explore such interfacial superconducting states. Dirac surface bands and possible topological superconductivity have been discussed[15, 16], while surface magnetism has also been observed[17-19], suggesting a setting in which superconductivity, topology, and magnetism may coexist. However, it remains difficult to determine how these properties are coupled at the interface and how their interplay manifests in transport. Moreover, freshly exposed surfaces of FTS are prone to degradation, making it difficult to preserve a well-defined interface in conventionally fabricated devices. Because FTS is a van der Waals (vdW) crystal, however, a pristine surface can be exposed by mechanical cleaving and immediately assembled into a junction without direct lithographic processing on the active interface. This provides a unique route to preserve the fragile surface states of FTS and to probe interfacial superconducting transport in a controlled geometry.

Here we exploit this van der Waals nature of FTS to realize Josephson junctions with a well-defined, freshly cleaved interface, using low-temperature micro-cleaving and stacking on a Peltier-cooled stage under an inert atmosphere. We observe nonreciprocal superconducting transport, including a zero-field JDE and field-dependent diode response, together with a hysteretic Fraunhofer pattern, magnetic-resonance-induced voltage features, and asymmetric half-integer Shapiro steps. These results suggest that the nonreciprocal transport of the FTS vdW Josephson junction is closely associated with emergent interfacial magnetism coupled with Josephson supercurrent. More broadly, our results establish vdW Josephson junctions as a platform for probing complex interfacial states in which magnetism, symmetry breaking, and superconductivity are intertwined.

**Interface-preserved FTS vdW Josephson junctions**

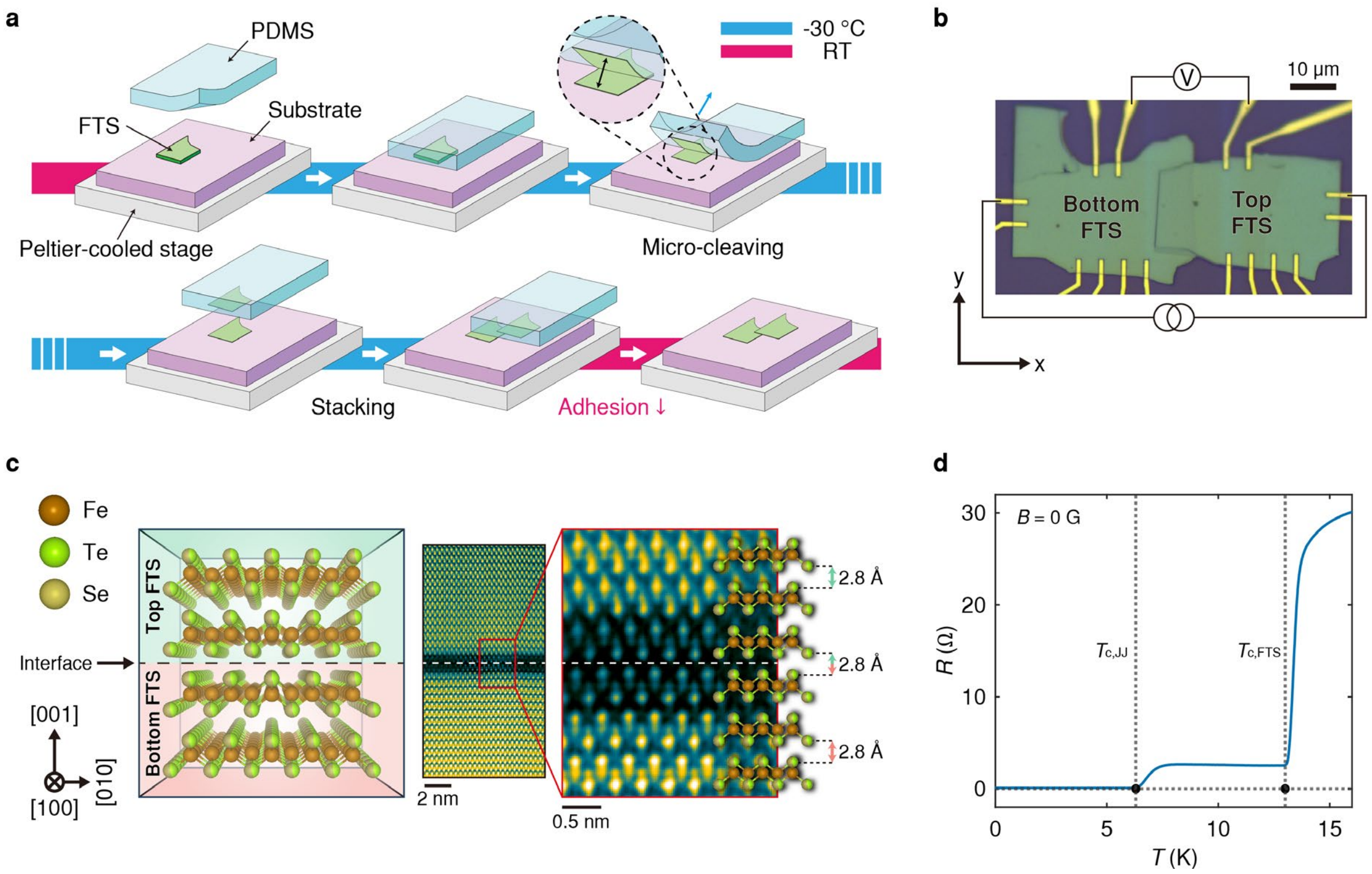

**Figure 1 | Fabrication and characterization of the $FeTe_{0.55}Se_{0.45}$ (FTS) van der Waals (vdW) Josephson junctions. a**, Schematic fabrication process of micro-cleaving and stacking technique on Peltier-cooled stage. **b**, Optical photograph of an FTS vdW Josephson junction composed of top and bottom FTS flakes with Cr/Au electrodes. **c**, Atomic configuration of FTS interface (left), cross-section scanning transmission electron microscope (STEM) image of a vdW Josephson junction (middle), and zoom-in image (right). Spheres denote atoms; Brown: Fe atoms, green: Te atoms, olive: Se atoms. The dashed line indicates the vdW interface. The atomic model and cross-sectional STEM images are shown along the $[100]_{\text{zone axis}}$. **d**, Temperature-dependent resistance $R(T)$ of the FTS vdW Josephson junction with a 1 $\mu A$ source current at zero-field, showing critical temperature of the FTS $T_{c,FTS}$, and that of vdW Josephson junction (JJ) $T_{c,JJ}$.

Surface integrity is a key constraint in fabricating vdW Josephson junctions from FTS, because air- or heat-induced unintended surface degradation occurs primarily at freshly exposed crystal surfaces.

While a quantitative, materials-specific assessment of FTS remains limited, related iron chalcogenides such as FeSe are known to exhibit pronounced surface evolution upon ambient and thermal exposure[20]. Consistently, we also observed noticeable surface changes in FTS under air and heat exposure, suggesting its susceptibility to degradation.

As shown in Fig. 1a, to suppress interfacial degradation during device fabrication, we adopted a low-temperature micro-cleaving and stacking procedure in an inert atmosphere, based on a strategy previously reported for BSCCO-based vdW Josephson junctions[21]. A single FTS crystal exfoliated on a substrate was split into two flakes using a polydimethylsiloxane (PDMS) stamp on a stage Peltier-cooled at $-30$ ℃ in an argon atmosphere; this process is referred to as micro-cleaving. After micro-cleaving, the upper flake was laterally shifted and placed back onto the lower flake; this process is referred to as stacking. The key idea is that the freshly cleaved surfaces were immediately restacked to form the vdW junction. Finally, Cr/Au electrodes are patterned and evaporated, followed by the deposition of a thin $Al_2O_3$ capping layer to protect the device during subsequent processing, as shown in Fig. 1b (see Methods for details of the fabrication process).

Figure 1c shows the structural integrity of the stacked FTS junction interface along $[100]_{\text{zone axis}}$, as examined by a high-angle annular dark-field scanning transmission electron microscopy (HAADF-STEM). The top and bottom FTS are clearly resolved across the vdW interface, indicated by the dashed lines. Because the top flake was stacked without intentional rotation, the two flakes remain crystallographically aligned across the junction, preserving the in-plane [010] and the out-of-plane [001] stacking direction. Moreover, the vdW interfacial spacing is approximately 2.8 Å, indicating that the junction interface preserves the native layer spacing of FTS with minimal structural discontinuity.

Transport measurements provide direct evidence for Josephson junction formation. As shown in Fig. 1d, after zero-field cooling, the temperature-dependent resistance $R(T)$ measured exhibits a two-step transition: near the critical temperature of FTS[22], $T_{\mathrm{c,FTS}}$ ~ 13.5 K, the FTS flakes become superconducting, and their resistance decreases, while the junction remains resistive. Below the critical temperature of Josephson junction $T_{\mathrm{c,JJ}}$ ~ 6 K, the device reaches a zero-resistance state, consistent with the onset of Josephson coupling across the vdW interface.

## Magnetization-dependent Josephson diode effect

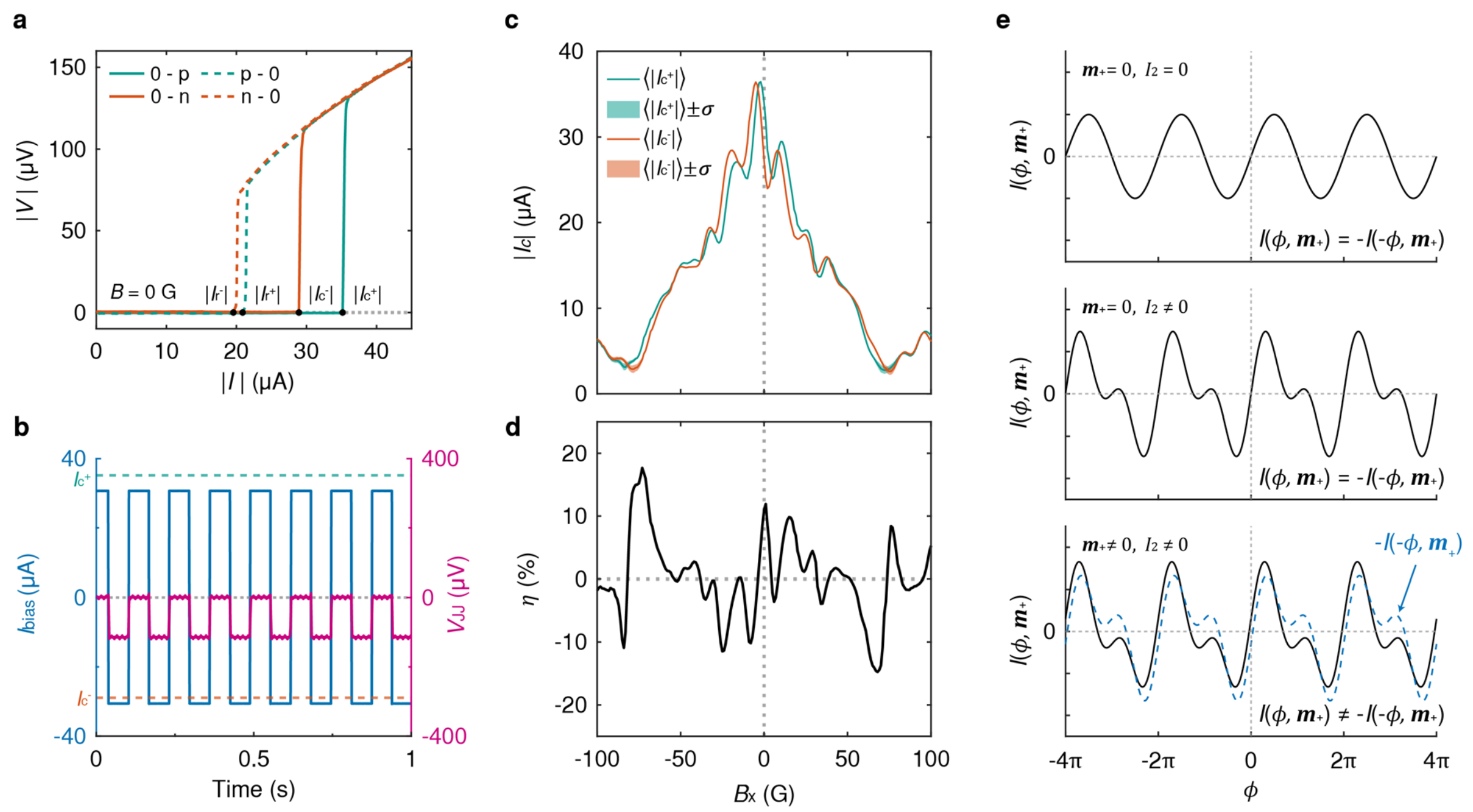

**Figure 2 | Josephson diode effect and magnetization-dependent current-phase relation (CPR) in an FTS vdW Josephson junction. a**, Current-voltage ($I-V$) characteristics of the FTS vdW Josephson junction at zero magnetic field. Solid traces show the switching branches, defined sweeping the current from zero to positive (0-p, teal) and from zero to negative (0-n, orange), from which the switching current $|I_c^+|$ and $|I_c^-|$ are extracted. Dashed traces show the retrapping branches defined by sweeping the current from positive back to zero (p-0, teal) and from negative back to zero (n-0, orange), from which $|I_r^+|$ and $|I_r^-|$ are determined. Black markers indicate $|I_c^\pm|$ and $|I_r^\pm|$. **b**, Time trace of the junction voltage $V_{JJ}$ (right axis, magenta) under square-wave current bias $I_{bias}$ (left axis, blue) with amplitude $32.5\ \mu A$ at $7.777\ Hz$. Dashed horizontal lines indicate the switching currents $I_c^+$ (teal) and $I_c^-$ (orange). **c**, Magnetic field dependence of the mean switching currents $\langle|I_c^+|\rangle$ (teal) and $\langle|I_c^-|\rangle$ (orange), extracted from repeated $I-V$ measurements $\langle|I_c^+|\rangle$ as a function of $B_x$. Shaded bands indicate standard deviation $\pm\sigma$ over 100 sweeps. **d**, Diode efficiency $\eta$ as a function of $B_x$, calculated from (**c**). **e**, Schematic magnetization-dependent CPR, $I(\phi, \boldsymbol{m}_+)$, where $\boldsymbol{m}_+$ denotes the effective aligned magnetization, and $I_2$ denotes the second-harmonic component.

The current-voltage ($I-V$) characteristics of the Josephson junction probe the transport behaviour of the Josephson current across the vdW interface and directly reflect the interfacial condition. After zero-field cooling to 20 mK, we measured the transport response of the same junction as a function of current sweep direction, as shown in Fig. 2a. To account for the low-temperature hysteresis, the current sweep is divided into four branches: 0–p, p–0, 0–n, and n–0. The 0–p and 0–n branches define the switching currents $|I_c^+|$ and $|I_c^-|$, while the p–0 and n–0 branches define the retrapping currents $|I_r^+|$ and $|I_r^-|$. The switching currents are asymmetric with respect to the current direction, with $|I_c^+| = 35.2\ \mu A$ and $|I_c^-| = 28.9\ \mu A$. The retrapping currents can also exhibit directional asymmetry through differences between $|I_r^+|$ and $|I_r^-|$. The observation of direction-dependent switching currents in the absence of an external magnetic field provides experimental evidence for nonreciprocal transport[23].

Figure 2b shows a half-wave rectification response under a square-wave current bias arising from the asymmetric switching currents. A square-wave current of 32.5 μA at 7.777 Hz is applied such that $|I_c^+| > 32.5\ \mu A > |I_c^-|$. As a result, the junction switches to the resistive state only during the negative half-cycle, producing a finite voltage, while remaining superconducting during the positive half-cycle. The reproducible response confirms that the asymmetry in $|I_c^+|$ and $|I_c^-|$ is not due to stochastic fluctuations (see Supplementary Note 2 for durability). From the measured switching currents, we obtain $\Delta I_c = |I_c^+| - |I_c^-| = 6.3\ \mu A$ and a diode efficiency of $\eta = \frac{|I_c^+|-|I_c^-|}{|I_c^+|+|I_c^-|} = 9.8\ \%$. Because bulk FTS crystallizes in a centrosymmetric P4/nmm structure[15], the observed zero-field diode response is unlikely to originate from the bulk crystal alone. Instead, this effect arises from inversion-symmetry-breaking and time-reversal-symmetry-breaking at the vdW interface.

The field-dependent evolution of the diode response was further examined under an in-plane magnetic field. Figure 2c plots mean switching current $\langle|I_c^+|\rangle$ (teal) and $\langle|I_c^-|\rangle$ (orange), with standard deviation $\pm\sigma$ bands obtained from 100 repeated $I-V$ sweeps under an in-plane magnetic field $B_x$ (see Methods). Both switching currents $\langle|I_c^+|\rangle$ and $\langle|I_c^-|\rangle$ show similar overall patterns symmetric around $B_x = -8\ G$, their maxima occur at slightly different field, around $B_x = -6\ G$ and $-10\ G$, respectively. This offset leads to field-dependent nonreciprocity between $\langle|I_c^+|\rangle$ and $\langle|I_c^-|\rangle$, quantified by the diode efficiency $\eta$ in Fig. 2d. Consistently, $\eta$ shows an antisymmetric dependence about $B_x = -8\ G$, varying from about -20 % to +20 % over the measured range.

This behaviour can be understood as an effective flux shift arising from an Edelstein-type magnetoelectric coupling at the inversion-broken vdW interface[24], where the Josephson current couples to the effective aligned magnetization. The resulting magnetization-dependent phase shift produces an offset in $B_{\mathrm{x}}$ between the Fraunhofer envelopes of $\langle|I_c^+|\rangle$ and $\langle|I_c^-|\rangle$, giving rise to the field-dependent diode efficiency. A faster oscillatory component with a periodicity of approximately 11 G is superimposed on the main Fraunhofer envelope. This feature may arise from an additional transport channel operating in parallel with the dominant Josephson current across the vdW interface, or from other spatial interference effects[25, 26]. As its microscopic origin is not directly relevant to the JDE that is the focus of this work, we do not discuss this component further in the main text.

To identify the observed nonreciprocal switching as a JDE, the asymmetry must originate from the Josephson phase response. In a conventional Josephson junction, the CPR depends only on the phase difference $\phi$, and unequal positive and negative critical currents require the breaking of the odd symmetry, $I(\phi) \neq -I(-\phi)$. Surface magnetism has previously been reported in FTS[17], and the observation of a zero-field diode response in the present vdW junction motivates consideration of its role. Following previous theoretical work[27], SOC at an inversion-symmetry-breaking interface mediates a coupling between the interfacial magnetization and the Josephson current, leading to a magnetization-dependent CPR,

$$I(\phi, \mathbf{m}_+) = I_1 \sin(\phi - \boldsymbol{\alpha} \cdot \mathbf{m}_+) + I_2 \sin 2\phi. \quad (1)$$

Here, $\boldsymbol{m}_t$ and $\boldsymbol{m}_b$ denote the normalized surface magnetizations of the top and bottom FTS layers, respectively, and $\boldsymbol{m}_+ = (\boldsymbol{m}_t + \boldsymbol{m}_b)/2$ represents their effective aligned magnetization. The vector $\boldsymbol{\alpha}$ characterizes the SOC arising from inversion-symmetry breaking, while $I_1$ and $I_2$ denote the first- and second-harmonic components of the CPR. Intuitively, at an interface with Rashba SOC, Cooper pairs acquire an additional geometric phase that depends on the orientation of the local magnetization, causing the Josephson current to depend not only on the superconducting phase difference but also on the interfacial magnetization (see Supplementary Note 1 for theoretical details).

Figure 2e illustrates the symmetry properties of Eq. (1). For $\boldsymbol{m}_+ = 0$, the CPR retains odd symmetry even in the presence of a finite second harmonic. In contrast, when both $\boldsymbol{m}_+ \neq 0$ and $I_2 \neq 0$, the phase-shifted first harmonic and the second harmonic act together to break this symmetry, resulting in unequal positive and negative critical currents and thereby enabling the JDE[9, 28].

## Hysteretic Josephson interference and magnetic resonance

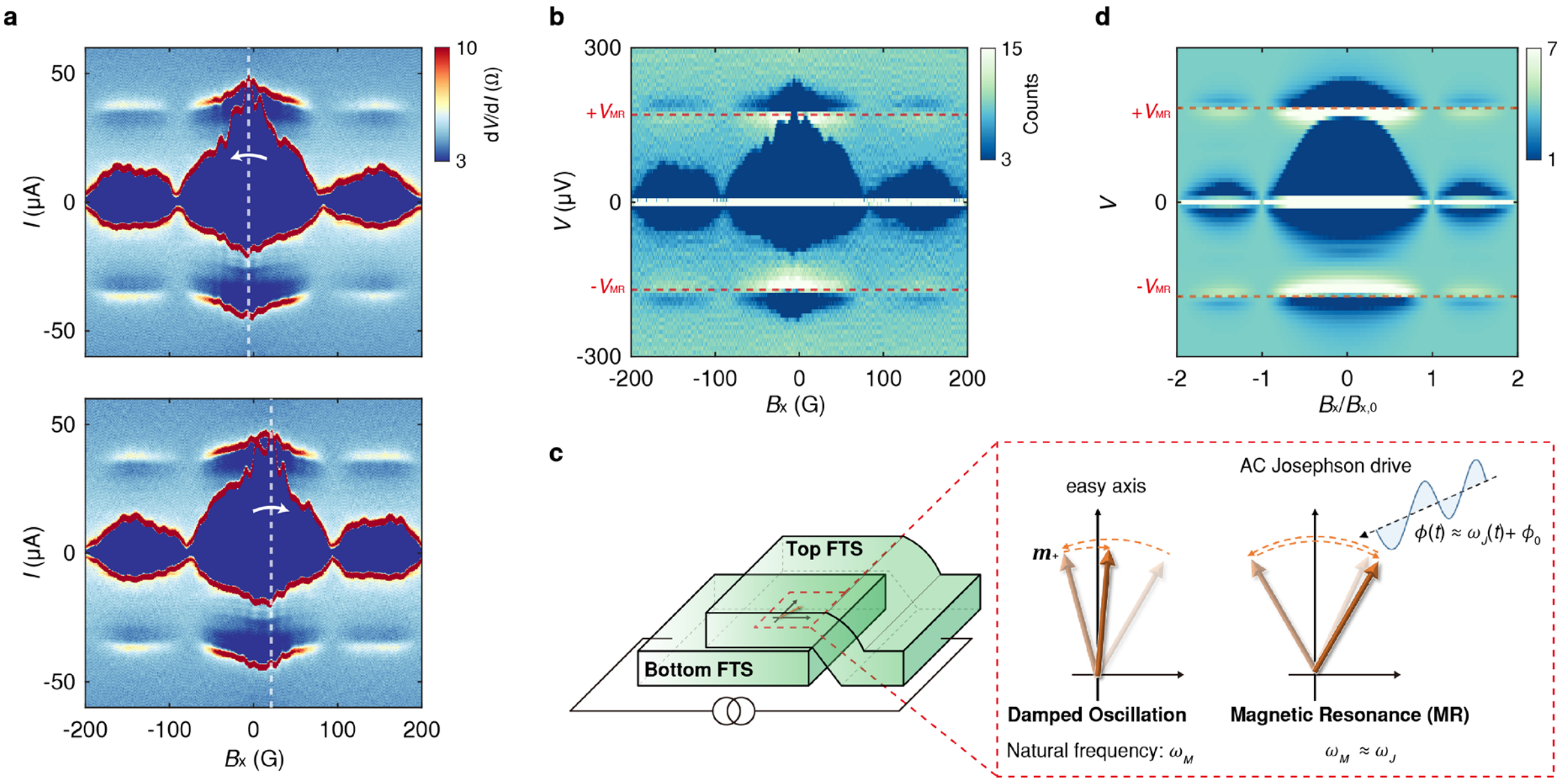

**Figure 3 | Hysteretic Fraunhofer patterns and magnetic resonance (MR)-induced voltage steps.** **a**, Differential conductance ($\mathrm{d}V/\mathrm{d}I$) map shows hysteretic Fraunhofer patterns of the FTS vdW Josephson junction. Top and bottom panels correspond to positive-to-negative and negative-to-positive field sweeps, respectively. White dashed lines indicate the main-lobe maxima at $B_\mathrm{x} = -6\ \mathrm{G}$ and $+22\ \mathrm{G}$. **b**, Bin-count map constructed from the $I-V$ data in the upper panel of (**a**) plotted versus $B_\mathrm{x}$ and $V$. Red dashed lines mark MR-induced voltage steps at $V_\mathrm{MR} = \pm 170\ \mu\mathrm{V}$. **c**, Schematic of MR induced by Josephson current. The effective aligned magnetization $\boldsymbol{m}_+$ undergoes damped oscillation with natural frequency $\omega_M$. When the Josephson frequency $\omega_J$ approaches $\omega_M$, resonant coupling drives magnetic resonance of $\boldsymbol{m}_+$. **d**, Bin-count map simulated using coupled RCSJ + LLG equations (see Supplementary Note 3 for model and parameters). $B_{x,0}$ is the period of Fraunhofer pattern.

The dependence of the Fraunhofer pattern on the magnetic-field sweep direction provides a means to probe the interfacial magnetic state associated with the magnetization-dependent CPR of Eq. (1). Figure 3a shows differential-resistance ($\mathrm{d}V/\mathrm{d}I$) maps measured for opposite $B_\mathrm{x}$ sweep directions. The main-lobe maxima exhibit clear sweep-direction dependence, revealing a hysteretic response that is absent in conventional non-magnetic Josephson junctions. Such behavior is consistent with a field-history-dependent magnetic state at the junction interface, analogous to hysteretic Fraunhofer responses reported in superconductor–ferromagnet–superconductor (SFS) Josephson junctions[29]. The observation therefore supports the presence of interfacial magnetism in the present vdW junction. Similar hysteretic Fraunhofer patterns have also been reported in vdW Josephson junctions formed by stacking independently exfoliated FTS flakes[19].

However, the presence of magnetization within the junction does not by itself establish its coupling to the Josephson current. Evidence for such coupling is provided by a distinct voltage-step feature observed in the finite-voltage regime. Re-examination of the $\mathrm{d}V/\mathrm{d}I$ map in Fig. 3a reveals an additional low-differential-resistance region above the switching current. Although its position in bias current varies with magnetic field, the voltage-axis representation of the same data in Fig. 3b shows that this feature remains concentrated around a nearly field-independent characteristic voltage, $V_{\mathrm{MR}} \cong 170\ \mu\mathrm{V}$. Notably, the feature appears within the Fraunhofer lobes and is suppressed near the nodes. This behavior distinguishes it from conventional Fiske resonances[29-32], which originate from geometric cavity modes of the Josephson junction and are not expected to track the Josephson current distribution in this manner. A detailed comparison is provided in the Supplementary Note 4.

To understand the origin of this characteristic voltage, we consider the coupling between the Josephson current and the effective aligned magnetization, as schematically illustrated in Fig. 3c. In the finite-voltage state, the superconducting phase evolves at the Josephson angular frequency, $\omega_J = 2eV/\hbar$, causing the Josephson current to oscillate at the same frequency. When $\omega_J$ approaches the natural frequency of the interfacial magnetization, $\omega_M$, the oscillating Josephson current can resonantly drive magnetization dynamics, giving rise to magnetic resonance (MR)[33]. This resonance appears as voltage steps in the $I-V$ response. Numerical simulations combining Landau–Lifshitz–Gilbert (LLG) dynamics with a resistively and capacitively shunted junction (RCSJ) model reproduce the observed behavior, including voltage steps at $\pm V_{\mathrm{MR}}$ (Fig. 3d; see Supplementary Note 4 for details). Using the Josephson relation, $V_{\mathrm{MR}} \cong 170\ \mu\mathrm{V}$ corresponds to a resonance frequency of approximately 82 GHz. This value is higher than typical ferromagnetic resonance frequencies (~10 GHz)[34, 35], but lower than conventional antiferromagnetic resonance frequencies (>100 GHz)[36, 37]. Such an intermediate energy scale is naturally explained by our model, in which predominantly in-plane magnetic moments, ferromagnetically aligned along the crystalline axes within each layer, are antiferromagnetically coupled between the top and bottom layers via a relatively weak interlayer exchange interaction, thereby producing a softened antiferromagnetic resonance at frequencies below those of conventional antiferromagnets.

## Higher-harmonic and phase-shifted Josephson dynamics

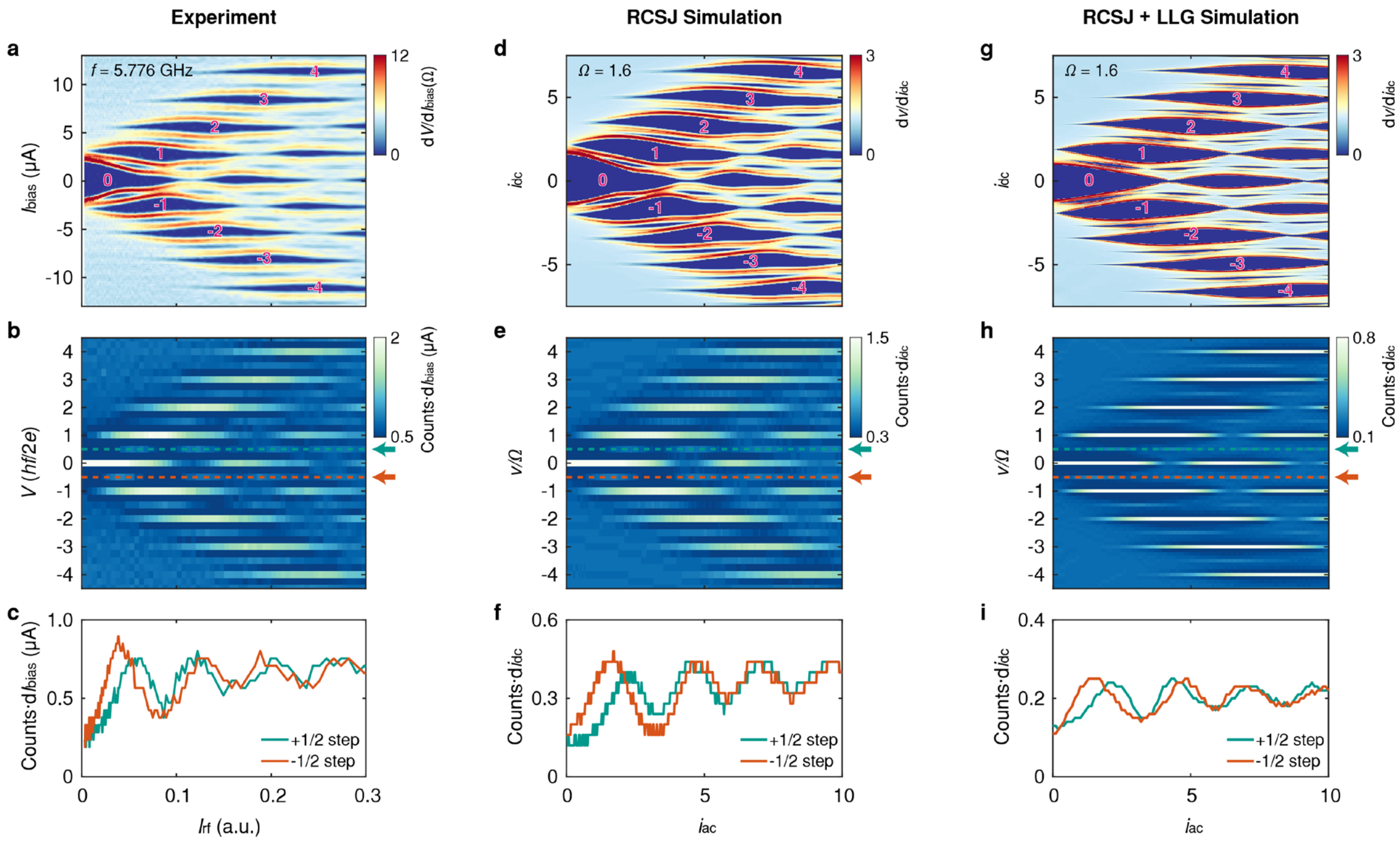

**Figure 4 | Shapiro-step response at $B_x = 75\ \mathrm{G}$ and numerical simulation. a**, Microwave response at $f = 5.776\ \mathrm{GHz}$: differential resistance $\mathrm{d}V/\mathrm{d}I$ as a function of dc bias current $I_{\mathrm{bias}}$ and radio-frequency (RF) drive amplitude $I_{\mathrm{rf}}$. Integer Shapiro steps ($n = 0, \pm1, \pm2, \ldots$) are labelled. **b**, Bin-count map constructed from (**a**), plotted versus $I_{\mathrm{rf}}$ and voltage normalized by $hf/2e$. Shapiro plateaus appear as maxima at fixed normalized voltages. Here, $\mathrm{d}I_{\mathrm{bias}}$ is current bias step size, so $\mathrm{Counts} \cdot \mathrm{d}I_{\mathrm{bias}}$ corresponds to the current width of the voltage plateau. Arrows indicate the half-integer positions $n = +1/2$ (teal) and $n = -1/2$ (orange). **c**, Line cuts of (**b**) at $n = +1/2$ (teal) and $n = -1/2$ (orange) as a function of $I_{\mathrm{rf}}$. **d–f**, Numerical RCSJ simulations with phase-only CPR (see Supplementary Note 6 for details); **d**, $\mathrm{d}v/\mathrm{d}i_{\mathrm{dc}}$ as a function of ac and dc current normalized by critical current, $i_{\mathrm{ac}}$ and $i_{\mathrm{dc}}$ at fixed dimensionless frequency $\Omega$, where $v$ is the dimensionless junction voltage. **e**, Bin-count map

from (**d**) plotted versus $i_{ac}$ and $v/\Omega$. **f**, Line cuts of (**e**) at $n = \pm 1/2$. **g–i**, Numerical RCSJ + LLG simulation (see Supplementary Note 6 for details), corresponding to the same quantities shown in (**d-f**).

Having established the presence of interfacial magnetism and its coupling to the Josephson current, we next investigate how the coupled dynamics of the magnetization and Josephson current are reflected in the Josephson phase dynamics. Direct reconstruction of the CPR is experimentally challenging, but Shapiro-step measurements provide a sensitive probe of higher-harmonic and phase-shifted CPR components[28]. We therefore performed Shapiro-step measurements at $B_x = 75$ G, near the first Fraunhofer node, where the JDE remains pronounced while the MR induced voltage feature is strongly suppressed. This choice minimizes the influence of MR related effects on the microwave response (see Supplementary Note 5 for Shapiro data at $B_x = 0$ G).

Figure 4a shows the microwave response measured at $f = 5.776$ GHz as a function of dc bias current $I_{bias}$ and radio-frequency (RF) drive amplitude $I_{rf}$. Integer Shapiro steps appear as regions of suppressed differential resistance $dV/dI$ at $V = nhf/2e$ with integer index $n = 0, \pm 1, \pm 2, \ldots$. Here, $h$ and $e$ represent Planck's constant and electron charge, respectively. In addition to these integer Shapiro steps, weaker half-integer Shapiro steps are observed between adjacent integer ones. To quantify these features, we construct a bin-count map of the voltage-normalized data as shown in Fig. 4b (see Methods). The voltage plateaus appear as enhanced bin counts at fixed values of $V/(hf/2e)$ at integer $n$ and weaker but reproducible enhanced counts at $n = \pm 1/2$. Line cuts at $n = +1/2$ and $-1/2$ (Fig. 4c) show distinct oscillatory dependence on $I_{rf}$, revealing an asymmetry between the positive and negative half integer steps.

As a phenomenological first step, we analyze the Shapiro response using a RCSJ model with an extended CPR[28] that contains the symmetry-allowed components implied by Eq. (1). Numerical simulations based on the RCSJ model with phase-only CPR reproduce the main features of the experiment, including integer steps, half-integer steps, and the asymmetric evolution of the $n = +1/2$ and $-1/2$ steps (Fig. 4d-f). This shows that a finite second-harmonic component is required to reproduce the half-integer Shapiro steps, while an additional phase-shift component is needed to capture the asymmetry between the $\pm 1/2$ steps (see Supplementary Note 6 for the detailed simulation procedure and CPR-component analysis).

A quantitative description of the experimentally observed phenomenology requires a sizable second-harmonic contribution ($I_2/I_1 = 0.7$). This value exceeds the second harmonic typically expected from highly transparent Josephson junctions ($I_2/I_1 < 0.4$)[38, 39]. Although a large second harmonic contribution can also appear in conventional SFS junction near a $0-\pi$ transition[40, 41], no signature of such a $0-\pi$ transition is observed in the present junction. This suggests that the observed half-integer steps involve an additional contribution beyond the intrinsic $I_2$ component. The origin of the enhanced second-harmonic contribution can be traced to the feedback between the Josephson dynamics and magnetization dynamics. In coupled RCSJ + LLG simulations, this feedback generates a significantly enhanced effective second-harmonic response even for a modest intrinsic value, $I_2/I_1 = 0.1$. The same framework also shows that a finite magnetization-induced phase shift $\boldsymbol{\alpha} \cdot \mathbf{m}_+$ accounts for the asymmetric evolution of the $\pm 1/2$ steps (see Supplementary Note 6 for details). These results suggest that magnetization–supercurrent coupling provides a unified mechanism for the enhanced second-harmonic response and finite phase shift required to account for the observed Shapiro response and the zero-field JDE.

## Discussion

In this work, we realized high-quality FTS vdW Josephson junctions using a low-temperature micro-cleaving and stacking process and identified a zero-field JDE. The hysteretic Fraunhofer pattern, MR-induced voltage features, and asymmetric half-integer Shapiro response indicate that the nonreciprocal Josephson transport is linked to coupling between the Josephson current and interfacial magnetism mediated by inversion-broken SOC. Beyond establishing FTS vdW junctions as a platform for anomalous Josephson transport—with the potential for reciprocal control between the Josephson current and interfacial magnetization—our observations also provide insight into the magnetic structure of the FTS interfaces. In particular, they suggest antiferromagnetic alignment between the two interfacial magnetic layers, while preserving ferromagnetic order within each individual layer.

More broadly, our results show that the microscopic origin of the JDE is not determined solely by the coexistence of broken time-reversal and inversion symmetries, but can depend on how interfacial magnetization couples to the Josephson current. This magnetization-Josephson current coupling provides a useful transport framework for understanding and designing nonreciprocal superconducting devices and may be extended to other magnetic Josephson junctions, including SFS junctions with strong interfacial SOC[42] and junctions based on materials where magnetism or magnetic fluctuations are intertwined with superconductivity, such as $UPd_2Al_3$[43, 44] and $UGe_2$[45] and $UTe_2$[46, 47].

## Methods

### Synthesis of $FeTe_{0.55}Se_{0.45}$ crystal

The single crystals were grown by horizontal versions of the Bridgman technique. As a start, the raw materials of 99.999 % Te, 99.999 % Se and 99.98 % Fe were weighed and mixed with the desired molar ratio, and then sealed into a high-purity (99.995 %) evacuated quartz tube. Since the tube often cracked upon cooling, it was sealed inside another larger evacuated tube. The doubly-sealed quartz tubes with the materials were then put into a box furnace horizontally crystal growth process. The quartz tube was heated up to 660 ℃ and keep for 12 hours; then heated up to 900 ℃ for 12 hours, then heated up to 1050 ℃ for 12 hours. The quartz tube was rocked for 10 hours at 1050 ℃ before the quartz tube was cooled to 950 ℃ at 0.5 hour. The tube was cooled at 0.5 ℃ per hours to 250 ℃ and then cooled to room temperature, and then shut down the furnace and cooled to room temperature. In the furnace, there was a small temperature gradient from one end to the other (e. g., at 850 ℃, $\Delta T$/distance ≈ 5 ℃/cm) such that the melted liquid crystallized unidirectionally. By using this method, large-size (> 10 g) high-quality single crystals can be obtained for $FeTe_{0.55}Se_{0.45}$. The crystals have nice mirror-like cleavage surfaces, which are the a-b planes. Many measurements such as resistivity, magnetization, x-ray diffraction (XRD) and neutron diffraction have been carried out on the crystals and confirm its quality and superconductivity.

### Device fabrication

The device was fabricated as illustrated in Fig. 1a of the main text to minimize thermal and air exposure during processing. All micro-cleaving and stacking procedures were performed on a home-built Peltier-cooled transfer stage inside an Ar-filled glove box, where the $H_2O$ and $O_2$ levels were maintained below 1 ppm. Immediately after stacking, the sample was transferred to the metal deposition chamber under an inert atmosphere and capped with a 2.5 nm-thick $Al_2O_3$ layer. The electrodes were subsequently defined by electron beam lithography, and the exposed contact regions were cleaned by Ar ion milling, followed by in-situ deposition of 6/100 nm-thick Cr/Au electrodes.

**Interface characterization**

Cross-sectional STEM specimens were prepared using a focused ion beam (FIB) system (Helios NanoLab G3 CX, FEI) to assess the degree of interfacial damage in the junction and to determine the size of the vdW gap. The cross-sectional STEM measurements were carried out at the Materials Imaging & Analysis Center of POSTECH using an ARM-200F microscope (JEOL Ltd., Japan) equipped with a probe corrector (ASCOR, CEOS GmbH, Germany). Images were acquired at an acceleration voltage of 80 kV using a high-angle annular dark-field detector with inner and outer collection angles of 54 and 216 mrad, respectively, together with a 40 μm condenser aperture.

**Transport measurements**

Transport measurements of the Josephson junction devices were performed in an Oxford ProteoxMX300 [6,1,1] vector magnet dilution refrigerator with a base temperature of 20 mK. The $R(T)$ characteristics of the Josephson junctions were measured using an SR830 lock-in amplifier. The $I-V$ curves, Fraunhofer pattern, and Shapiro step data were obtained by applying a DC current bias using a GS610 connected in series with a 100 $\mathrm{k\Omega}$ standard resistor, while the voltage was measured with a Keithley 2000 multimeter. For the Shapiro step measurements, microwave power in the range of -60~10 dBm was applied to the device using an 8340B synthesized sweeper. For the Josephson diode effect measurements, a triangular-current bias at 7.777 Hz was applied using a WF1947 multifunction generator, and the current and voltage were simultaneously recorded using an NI USB-6361. The stability of the JDE was examined using the same measurement configuration, except that the current bias waveform was changed to a square-wave.

**Extraction of switching currents $\langle|\boldsymbol{I_c^{\pm}}|\rangle$**

Switching currents were extracted from continuous $I-V$ data acquired under 100 cycles of triangular current bias at each magnetic-field value. The data were first divided into individual sweep branches triangular waveform, and positive and negative switching events were identified from the n-p and p-n branches, respectively. The switching current was defined as the bias current at the first point where the absolute junction voltage crossed upward from below a threshold of 10 $\mu\mathrm{V}$ to above it and remained above the threshold over approximately one-tenth of the same branch. This criterion was used to exclude unintended crossings caused by transient voltage fluctuations or noise, thereby selecting only stable switching events. Using this procedure, 100 values of $|I_c^+|$ and 100 values of $|I_c^-|$ were obtained at each magnetic-field value, from which the absolute mean values $\langle|I_c^+|\rangle$ and $\langle|I_c^-|\rangle$ and their standard deviations were calculated (see Supplementary Note 2).

**Analysis of Shapiro steps**

For the Shapiro-step analysis, the measured $I-V$ data were normalized by $hf/2e$ in voltage. The normalized voltage axis was binned into bins with a width of $1/4$, such that each bin covered a range of $\pm 1/8$ around its center. For the RCSJ + LLG simulations, a finer bin width of $1/8$ was used so that each bin covered $\pm 1/16$, allowing additional fractional features such as $1/6$ steps to be resolved. For each RF drive condition, the number of data points falling within each bin was multiplied by the current interval $\mathrm{d}I_{\mathrm{bias}}$ to obtain the current width of the plateau within the corresponding normalized voltage range. This quantity was used as the color-scale value in the bin-count map. Therefore, a larger color-scale value indicates a wider current plateau formed within the corresponding normalized voltage interval. The horizontal axis, $I_{\mathrm{rf}}$ (a.u.), denotes the relative RF drive amplitude defined from the applied microwave power $P$ (dBm) as $I_{\mathrm{rf}} = 10^{P/20}$.

**Acknowledgements**

This research was supported by National Research Foundation of Korea (NRF) (Nos. RS-2022-NR068223, RS-2024-00393599, RS-2024-00442710, RS-2024-00444725, RS-2025-02317602) and the ITRC program (IITP-2025-RS-2022-00164799) funded by the Ministry of Science and ICT (MSIT). Further support was received from Samsung Science and Technology Foundation (SSTF-BA2401-03). C.-E.A. and G.Y.C. are also financially supported by the NRF of Korea (Grants No. RS-2024-00410027, RS-2023-NR119931, RS-2024-00444725, RS-2023-00256050, IRS-2025-25453111, RS-2025-08542968, RS-2026-25479545) funded by MSIT, the Air Force Office of Scientific Research under Award No. FA23862514026. The work at BNL was supported by the US Department of Energy, office of Basic Energy Sciences, contract no. DOE-SC0012704.

**Author contributions**

G.-H.L. and S.-H.S. conceived the project. S.-H.S. fabricated the devices, performed the transport measurements, and analysed the data. C.-E.A. and G.Y.C. developed the theoretical model and analysed the data. C.-E.A. and T.O. carried out the theoretical calculations. G.G. grew and provided the FTS crystals. C.-W.C. and S.-Y.C. performed the STEM measurements and analysis. S.-H.S., G.-H.L., C.-E.A., and G.Y.C. wrote the manuscript. All authors discussed the results and contributed to the manuscript.

**Competing financial interests**

The authors declare no competing financial interests.

**Data availability**

The data that supports the findings of this study are available from the corresponding authors upon reasonable request.

**References**

1. Tokura Y, Nagaosa N. Nonreciprocal responses from non-centrosymmetric quantum materials. *Nat Commun* 2018, **9**(1)**:** 3740.
2. Rikken GLJA, Wyder P. Magnetoelectric Anisotropy in Diffusive Transport. *Phys Rev Lett* 2005, **94**(1)**:** 016601.
3. Ando F, Miyasaka Y, Li T, Ishizuka J, Arakawa T, Shiota Y*, et al.* Observation of superconducting diode effect. *Nature* 2020, **584**(7821)**:** 373-376.
4. Nadeem M, Fuhrer MS, Wang X. The superconducting diode effect. *Nature Reviews Physics* 2023, **5**(10)**:** 558-577.
5. Chen C-Z, He JJ, Ali MN, Lee G-H, Fong KC, Law KT. Asymmetric Josephson effect in inversion symmetry breaking topological materials. *Phys Rev B* 2018, **98**(7)**:** 075430.
6. Reinhardt S, Ascherl T, Costa A, Berger J, Gronin S, Gardner GC*, et al.* Link between supercurrent diode and anomalous Josephson effect revealed by gate-controlled interferometry. *Nat Commun* 2024, **15**(1)**:** 4413.
7. Pal B, Chakraborty A, Sivakumar PK, Davydova M, Gopi AK, Pandeya AK*, et al.* Josephson diode effect from Cooper pair momentum in a topological semimetal. *Nat Phys* 2022, **18**(10)**:** 1228-1233.
8. Baumgartner C, Fuchs L, Costa A, Reinhardt S, Gronin S, Gardner GC*, et al.* Supercurrent

rectification and magnetochiral effects in symmetric Josephson junctions. *Nat Nanotechnol* 2022, **17**(1)**:** 39-44.

9. Fominov YV, Mikhailov DS. Asymmetric higher-harmonic SQUID as a Josephson diode. *Phys Rev B* 2022, **106**(13)**:** 134514.

10. Turini B, Salimian S, Carrega M, Iorio A, Strambini E, Giazotto F*, et al.* Josephson Diode Effect in High-Mobility InSb Nanoflags. *Nano Lett* 2022, **22**(21)**:** 8502-8508.

11. Gupta M, Graziano GV, Pendharkar M, Dong JT, Dempsey CP, Palmstrøm C*, et al.* Gate-tunable superconducting diode effect in a three-terminal Josephson device. *Nat Commun* 2023, **14**(1)**:** 3078.

12. Trahms M, Melischek L, Steiner JF, Mahendru B, Tamir I, Bogdanoff N*, et al.* Diode effect in Josephson junctions with a single magnetic atom. *Nature* 2023, **615**(7953)**:** 628-633.

13. Narita H, Ishizuka J, Kawarazaki R, Kan D, Shiota Y, Moriyama T*, et al.* Field-free superconducting diode effect in noncentrosymmetric superconductor/ferromagnet multilayers. *Nat Nanotechnol* 2022, **17**(8)**:** 823-828.

14. Willsch D, Rieger D, Winkel P, Willsch M, Dickel C, Krause J*, et al.* Observation of Josephson harmonics in tunnel junctions. *Nat Phys* 2024, **20**(5)**:** 815-821.

15. Wang Z, Zhang P, Xu G, Zeng LK, Miao H, Xu X*, et al.* Topological nature of the $FeSe_{0.5}Te_{0.5}$ superconductor. *Phys Rev B* 2015, **92**(11)**:** 115119.

16. Zhang P, Yaji K, Hashimoto T, Ota Y, Kondo T, Okazaki K*, et al.* Observation of topological superconductivity on the surface of an iron-based superconductor. *Science* 2018, **360**(6385)**:** 182-186.

17. Zaki N, Gu G, Tsvelik A, Wu C, Johnson PD. Time-reversal symmetry breaking in the Fe-chalcogenide superconductors. *Proceedings of the National Academy of Sciences* 2021, **118**(3)**:** e2007241118.

18. Farhang C, Zaki N, Wang J, Gu G, Johnson PD, Xia J. Revealing the Origin of Time-Reversal Symmetry Breaking in Fe-Chalcogenide Superconductor $FeTe_{1-x}Se_x$. *Phys Rev Lett* 2023, **130**(4)**:** 046702.

19. Qiu G, Yang H-Y, Hu L, Zhang H, Chen C-Y, Lyu Y*, et al.* Emergent ferromagnetism with superconductivity in Fe(Te,Se) van der Waals Josephson junctions. *Nat Commun* 2023, **14**(1)**:** 6691.

20. Lanoël L, Rozas G, Bruchhausen AE, Amigó ML, Crivillero MVA, Hofer JA*, et al.* Effects of aging processes at the surface of the superconductor β–FeSe. *Phys Rev B* 2022, **106**(21)**:** 214507.

21. Zhao SYF, Cui X, Volkov PA, Yoo H, Lee S, Gardener JA*, et al.* Time-reversal symmetry breaking superconductivity between twisted cuprate superconductors. *Science* 2023, **382**(6677)**:** 1422-1427.

22. Yeh K-W, Huang T-W, Huang Y-I, Chen T-K, Hsu F-C, M. Wu P*, et al.* Tellurium substitution effect on superconductivity of the α-phase iron selenide. *Europhys Lett* 2008, **84**(3)**:** 37002.

23. Wu H, Wang Y, Xu Y, Sivakumar PK, Pasco C, Filippozzi U*, et al.* The field-free Josephson

diode in a van der Waals heterostructure. *Nature* 2022, **604**(7907)**:** 653-656.

24. Edelstein VM. Magnetoelectric Effect in Polar Superconductors. *Phys Rev Lett* 1995, **75**(10)**:** 2004-2007.
25. Gray MJ, Freudenstein J, Zhao SYF, O'Connor R, Jenkins S, Kumar N*, et al.* Evidence for Helical Hinge Zero Modes in an Fe-Based Superconductor. *Nano Lett* 2019, **19**(8)**:** 4890-4896.
26. Mayer W, Dartiailh MC, Yuan J, Wickramasinghe KS, Rossi E, Shabani J. Gate controlled anomalous phase shift in Al/InAs Josephson junctions. *Nat Commun* 2020, **11**(1)**:** 212.
27. Buzdin A. Direct Coupling Between Magnetism and Superconducting Current in the Josephson $\varphi_0$ Junction. *Phys Rev Lett* 2008, **101**(10)**:** 107005.
28. Volkov PA, Lantagne-Hurtubise É, Tummuru T, Plugge S, Pixley JH, Franz M. Josephson diode effects in twisted nodal superconductors. *Phys Rev B* 2024, **109**(9)**:** 094518.
29. Ai L, Zhang E, Yang J, Xie X, Yang Y, Jia Z*, et al.* Van der Waals ferromagnetic Josephson junctions. *Nat Commun* 2021, **12**(1)**:** 6580.
30. Coon DD, Fiske MD. Josephson ac and Step Structure in the Supercurrent Tunneling Characteristic. *Physical Review* 1965, **138**(3A)**:** A744-A746.
31. Fiske MD. Temperature and Magnetic Field Dependences of the Josephson Tunneling Current. *Rev Mod Phys* 1964, **36**(1)**:** 221-222.
32. Yabuki N, Moriya R, Arai M, Sata Y, Morikawa S, Masubuchi S*, et al.* Supercurrent in van der Waals Josephson junction. *Nat Commun* 2016, **7**(1)**:** 10616.
33. Shukrinov YM, Rahmonov IR, Sengupta K. Ferromagnetic resonance and magnetic precessions in $\varphi_0$ junctions. *Phys Rev B* 2019, **99**(22)**:** 224513.
34. Petković I, Aprili M, Barnes SE, Beuneu F, Maekawa S. Direct dynamical coupling of spin modes and singlet Josephson supercurrent in ferromagnetic Josephson junctions. *Phys Rev B* 2009, **80**(22)**:** 220502.
35. Barnes SE, Aprili M, Petković I, Maekawa S. Ferromagnetic resonance with a magnetic Josephson junction. *Superconductor Science and Technology* 2011, **24**(2)**:** 024020.
36. Wang Z, Kovalev S, Awari N, Chen M, Germanskiy S, Green B*, et al.* Magnetic field dependence of antiferromagnetic resonance in NiO. *Appl Phys Lett* 2018, **112**(25)**:** 252404.
37. Ouyang C, Wang X, Li Y. Chirality-selective easily adjustable spin current from uniaxial antiferromagnets. *Appl Phys Lett* 2021, **119**(14)**:** 142404.
38. Hart S, Cui Z, Ménard G, Deng M, Antipov AE, Lutchyn RM*, et al.* Current-phase relations of InAs nanowire Josephson junctions: From interacting to multimode regimes. *Phys Rev B* 2019, **100**(6)**:** 064523.
39. Zhang P, Zarassi A, Jarjat L, Van de Sande V, Pendharkar M, Lee JS*, et al.* Large second-order Josephson effect in planar superconductor-semiconductor junctions. *SciPost Physics* 2024, **16**(1)**:** 030.
40. Stoutimore MJA, Rossolenko AN, Bolginov VV, Oboznov VA, Rusanov AY, Baranov DS*, et al.* Second-Harmonic Current-Phase Relation in Josephson Junctions with Ferromagnetic Barriers. *Phys Rev Lett* 2018, **121**(17)**:** 177702.

41. Konschelle F, Cayssol J, Buzdin AI. Nonsinusoidal current-phase relation in strongly ferromagnetic and moderately disordered SFS junctions. *Phys Rev B* 2008, **78**(13)**:** 134505.
42. Satchell N, Birge NO. Supercurrent in ferromagnetic Josephson junctions with heavy metal interlayers. *Phys Rev B* 2018, **97**(21)**:** 214509.
43. Geibel C, Schank C, Thies S, Kitazawa H, Bredl CD, Böhm A*, et al.* Heavy-fermion superconductivity atTc=2K in the antiferromagnet UPd2Al3. *Zeitschrift für Physik B Condensed Matter* 1991, **84**(1)**:** 1-2.
44. He Y, Muirhead C, Bradshaw A, Abell J, Schank C, Geibel G*, et al.* Coherence of the superconducting wavefunction between the heavy-fermion superconductor UPd2Al3 and niobium. *Nature* 1992, **357**(6375)**:** 227-229.
45. Saxena SS, Agarwal P, Ahilan K, Grosche FM, Haselwimmer RKW, Steiner MJ*, et al.* Superconductivity on the border of itinerant-electron ferromagnetism in UGe2. *Nature* 2000, **406**(6796)**:** 587-592.
46. Sundar S, Gheidi S, Akintola K, Côté AM, Dunsiger SR, Ran S*, et al.* Coexistence of ferromagnetic fluctuations and superconductivity in the actinide superconductor $UTe_2$. *Phys Rev B* 2019, **100**(14)**:** 140502.
47. Li Z, Moir CM, McKee NJ, Lee-Wong E, Baumbach RE, Maple MB*, et al.* Observation of odd-parity superconductivity in UTe2. *Proceedings of the National Academy of Sciences* 2025, **122**(13)**:** e2419734122.

# Supplementary Information for

## Magnetization and Josephson current coupling drives the Josephson diode effect in $FeTe_{0.55}Se_{0.45}$ van der Waals junctions

Seung-Hyun Shin[1,†], Cheong-Eung Ahn[1,†], Chang-Won Choi[2, 6], Taekoo Oh[3], Si-Young Choi[2, 6, 7], Genda Gu[4], Gil Young Cho[5, 8,*] and Gil-Ho Lee[1,*]

[1] Department of Physics, POSTECH, Pohang, South Korea
[2] Department of Materials Science and Engineering, POSTECH, Pohang, South Korea
[3] Department of Physics, Soongsil University, Seoul, South Korea
[4] Condensed Matter Physics and Materials Science Department, Brookhaven National Laboratory, Upton, NY, USA
[5] Department of Physics, Korea Advanced Institute of Science and Technology, Daejeon, South Korea
[6] Center for Van der Waals Quantum Solids, Institute for Basic Science (IBS), Pohang, South Korea
[7] Department of Semiconductor Engineering, POSTECH, Pohang, South Korea
[8] Center for Artificial Low Dimensional Electronic Systems, Institute for Basic Science, Pohang 37673, Republic of Korea

†*These authors equally contributed.*

**Correspondence to Gil Young Cho (gilyoungcho@kaist.ac.kr) & Gil-Ho Lee (lghman@postech.ac.kr)*

## Contents

### Supplementary Note 1: Theoretical Model and Simulations

In this Supplementary Note, we present the theoretical model and corresponding numerical simulations underlying our analysis. Microscopically, inversion symmetry can be broken at the van der Waals interface by chemical-potential mismatch[1, 2] and by strain or structural relaxation[3-5]. This local inversion breaking may be further supplemented by the macroscopic inversion asymmetry of the device geometry itself. In particular, we assume inversion symmetry breaking primarily along the $x$-direction, giving a primarily $y$-aligned $\boldsymbol{\alpha}$ from Rashba spin–orbit coupling (SOC)[6]. This spin–orbit coupling enables the Josephson current flowing along the $z$-direction to couple to the in-plane interfacial magnetization.

The interfacial magnetization is expected to lie within the plane, further aligned along the crystalline axes, i.e., $x$-direction or $y$-direction, due to the in-plane magnetic anisotropy and the underlying $C_4$ point-group symmetry of the crystal. This is also consistent with the different critical current Fraunhofer patterns for $B_x$ and $B_y$ external fields. Other spin configurations, for example with moments aligned along the diagonal directions, are also possible, and lead to qualitatively similar results[7]. Furthermore, our experimental data are consistent with an antiferromagnetic exchange coupling between the magnetizations on the top and bottom surfaces, $\mathbf{m}_t$ and $\mathbf{m}_b$. Incorporating these ingredients, the full energy functional is given by

$$\frac{E_{\mathrm{mag}}(\mathbf{m}_t, \mathbf{m}_b, \phi)}{A} = \sum_{\ell=t,b} \left(\frac{1}{2}K_1 m_{\ell,z}^2 + \frac{1}{2}K_2 m_{\ell,x}^2 m_{\ell,y}^2\right) + \frac{1}{2}J_{\mathrm{ex}}\mathbf{m}_t \cdot \mathbf{m}_b - \frac{(\hbar/2e)I_1}{A}\cos(\phi - \boldsymbol{\alpha}\cdot\mathbf{m}_+), \qquad \text{(S1)}$$

where $A$ is the junction area, $K_1, K_2 > 0$ are the two magnetic anisotropies, and $J_{\mathrm{ex}} > 0$ is the interlayer antiferromagnetic exchange coupling. In the below, we will consider symmetric and anti-symmetric components of the magnetizations $\mathbf{m}_\pm = (\mathbf{m}_t \pm \mathbf{m}_b)/2$. The magnetization dynamics are then modeled using the Landau–Lifshitz–Gilbert (LLG) equation, including both the Zeeman coupling to external magnetic fields and the coupling to the Josephson current. Our parameters are the Gilbert damping factor $\zeta_M$, the normalized external magnetic field, $\mathbf{h}^{\mathrm{ext}} = \mathbf{B}^{\mathrm{ext}}/B^M$ where $B^M = \sqrt{(J_{\mathrm{ex}} + K_2)K_1}/M_S$ and $M_S$ is the saturation magnetization, the normalized coupling parameters,

$$\eta = \sqrt{\frac{K_1}{J_{\mathrm{ex}} + K_2}}, \qquad \kappa = \frac{K_2}{K_1}, \qquad \Gamma = \frac{1}{2}\frac{(\hbar/2e)I_1/A}{\sqrt{(J_{\mathrm{ex}} + K_2)K_1}},$$

and the magnetic characteristic frequency,

$$\omega_M = \frac{\gamma}{\mu_0 M_S}\sqrt{(J_{\mathrm{ex}} + K_2)K_1}.$$

Note that this natural frequency is set by $\omega_M \sim \sqrt{J_{\mathrm{ex}}K_1}$, taking the form of the antiferromagnetic natural frequency, as opposed to magnetic anisotropy driven ferromagnetic natural frequency, which scales with magnetic anisotropy alone[8].

The dynamics of the superconducting phase are modeled using the resistively and capacitively shunted junction (RCSJ) model.

$$\beta_J \frac{\mathrm{d}v}{\mathrm{d}(\omega_J t)} = -[i(\phi(t), \mathbf{m}_+(t)) - i_{\mathrm{net}}(t) + v], \qquad \frac{\mathrm{d}\phi}{\mathrm{d}(\omega_J t)} = v(t), \tag{S2}$$

where $v(t) = V(t)/RI_1$ is the normalized voltage, $i(\phi, \mathbf{m}_+) = I(\phi, \mathbf{m}_+)/I_1$ is the normalized current-phase relation, Eq. (1), $i_{\mathrm{net}}(t) = I_{\mathrm{net}}(t)/I_1$ is the normalized net current (DC only for the critical current and current-voltage characteristic, DC + AC for Shapiro steps), $i_2 = I_2/I_1$ is the normalized intrinsic second harmonic amplitude, $\omega_J = (2e/\hbar)I_1 R$ is the Josephson characteristic frequency, $\beta_J = \omega_J RC$ is the Stewart-McCumber parameter, representing the normalized inverse dissipation, and $R$, $C$ are the effective shunt resistance and capacitance of the Josephson junction.

When an external magnetic field is applied, the superconducting phase difference across the junction acquires spatial dependence. For a uniform external magnetic field $B_x$ along the $x$-direction, the superconducting phase can be written as

$$\phi(x, y; \Phi_x) \approx \tilde{\phi} + \frac{2\pi\Phi_x}{\Phi_0}\frac{y}{L_y},$$

where $\tilde{\phi} \equiv \phi(y = 0)$ denotes the reduced superconducting phase difference. Here, we approximate the junction geometry as a rectangular region of size $L_x \times L_y$, $\Phi_x = B_x(d_{\mathrm{eff}} L_y)$ is the magnetic flux through the junction, and $\Phi_0$ denotes the superconducting flux quantum. Based on this expression, we integrate the current-phase relation over the junction area to find the total current

$$\tilde{I}(\tilde{\phi}, \mathbf{m}_+; \Phi_x) = I_1 \frac{\sin(\pi\Phi_x/\Phi_0)}{\pi\Phi_x/\Phi_0} \sin(\tilde{\phi} - \boldsymbol{\alpha} \cdot \mathbf{m}_+) + I_2 \frac{\sin(2\pi\Phi_x/\Phi_0)}{2\pi\Phi_x/\Phi_0} \sin 2\tilde{\phi}, \tag{S3}$$

which replace the CPR Eq. (1) in Eq. (S2).

In this framework, one can further incorporate the feedback of the magnetization on the superconducting phase gradient. The full derivation and microscopic details are provided in a separate theoretical work[7], while we provide brief comments here where necessary.

In our simulations, we have used the fixed parameters,

$$\zeta_M = 5.0 \times 10^{-2}, \qquad \eta = 3.0 \times 10^{-1}, \qquad \kappa = 0.30, \qquad B^M = 2000 \ \mathrm{G}, \tag{S4}$$

while $\omega_M/\omega_J$, $\boldsymbol{\alpha}$, $\Gamma$, $\beta_J$, and $i_2$ have been adjusted to account for interference effects. These are listed in Supplementary Notes 3, 4, and 6.

## Supplementary Note 2: Additional Details of Critical Current Measurements

### A. Durability of the Josephson Diode Response

To evaluate the durability of the FTS Josephson diode response, we examined both its stability under repeated square-wave operation and its evolution with temperature. In Fig. 2 of the main text, $|I_c^+|$ and $|I_c^-|$ have different values at zero magnetic field, resulting in half-wave rectification under square-wave current bias, where a voltage response appears only for one current direction. However, it is necessary to verify whether this diode response indeed originates from the nonreciprocal nature of the Josephson junction, rather than from stochastic switching, transient instability, noise, or a temporary change in the junction state. Therefore, in Supplementary Fig. 1, we performed repeated square-wave measurements under the same bias condition to examine the long-term stability of the JDE response.

Supplementary Fig. 1a shows the junction voltage response when a square-wave current bias was applied for 5,000 cycles under the same condition as in Fig. 2b of the main text. The voltage pulses repeatedly appear only for one current direction (p-n), while the junction remains in the superconducting state for the opposite direction (n-p). The fact that this half-wave voltage response remains stable over thousands of repeated cycles indicates that the observed rectification does not arise from accidental switching events or transient noise. Rather, it shows that the FTS Josephson diode remains stable during repeated operation.

Supplementary Fig. 1b shows the magnetic-field dependence of the diode efficiency $\eta$ and its temperature evolution. At low temperature, $\eta$ exhibits an oscillatory dependence on magnetic field, reflecting the field-dependent imbalance between $|I_c^+|$ and $|I_c^-|$. As the temperature increases, the oscillation amplitude of $\eta$ is progressively suppressed, indicating that the nonreciprocal switching response becomes weaker at elevated temperatures. This temperature-dependent suppression is consistent with the diode response being associated with superconducting Josephson transport, rather than normal-state rectification or contact-related artifacts.

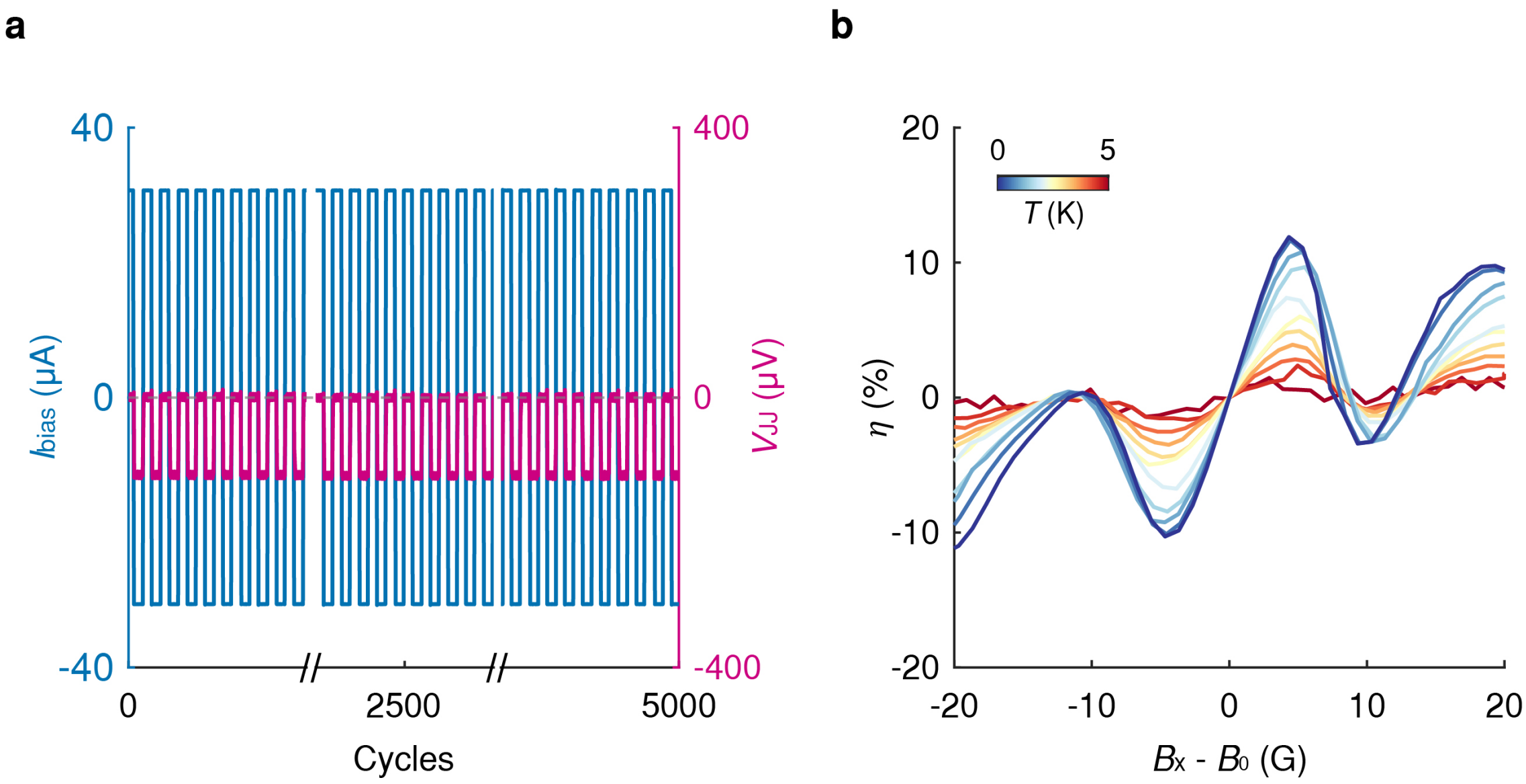

**Supplementary Fig. 1| Durability of the Josephson diode response and temperature evolution of the diode efficiency. a**, Square-wave current bias $I_{\mathrm{bias}}$ (left axis, blue) and the corresponding junction voltage $V_{\mathrm{JJ}}$ (right axis, magenta) measured over 5,000 cycles under the same conditions as in Fig. 2b. **b**, Magnetic-field dependence of the diode efficiency $\eta$ and its temperature evolution measured over the range $0{\sim}5$ K. The magnetic field is plotted relative to $B_0$, the symmetric field at which $\eta = 0$.

### B. Extraction of Switching Currents $\langle|I_c^{\pm}|\rangle$ (in Methods)

Supplementary Fig. 2 illustrates the procedure used to extract the switching currents from repeated triangular current bias measurements. The voltage response was separated according to the current sweep direction, and the switching events in the n-p and p-n branches were used to determine $|I_c^+|$ and $|I_c^-|$, respectively. Using this procedure, 100 values of $|I_c^+|$ and $|I_c^-|$ were obtained at each magnetic field value. Supplementary Fig. 2b shows representative histograms of the extracted switching currents, confirming that the difference between $|I_c^+|$ and $|I_c^-|$ is statistically reproducible rather than arising from a single switching event. The mean values and standard deviations obtained from these distributions were used for the Fraunhofer pattern and diode efficiency analyses in the main text.

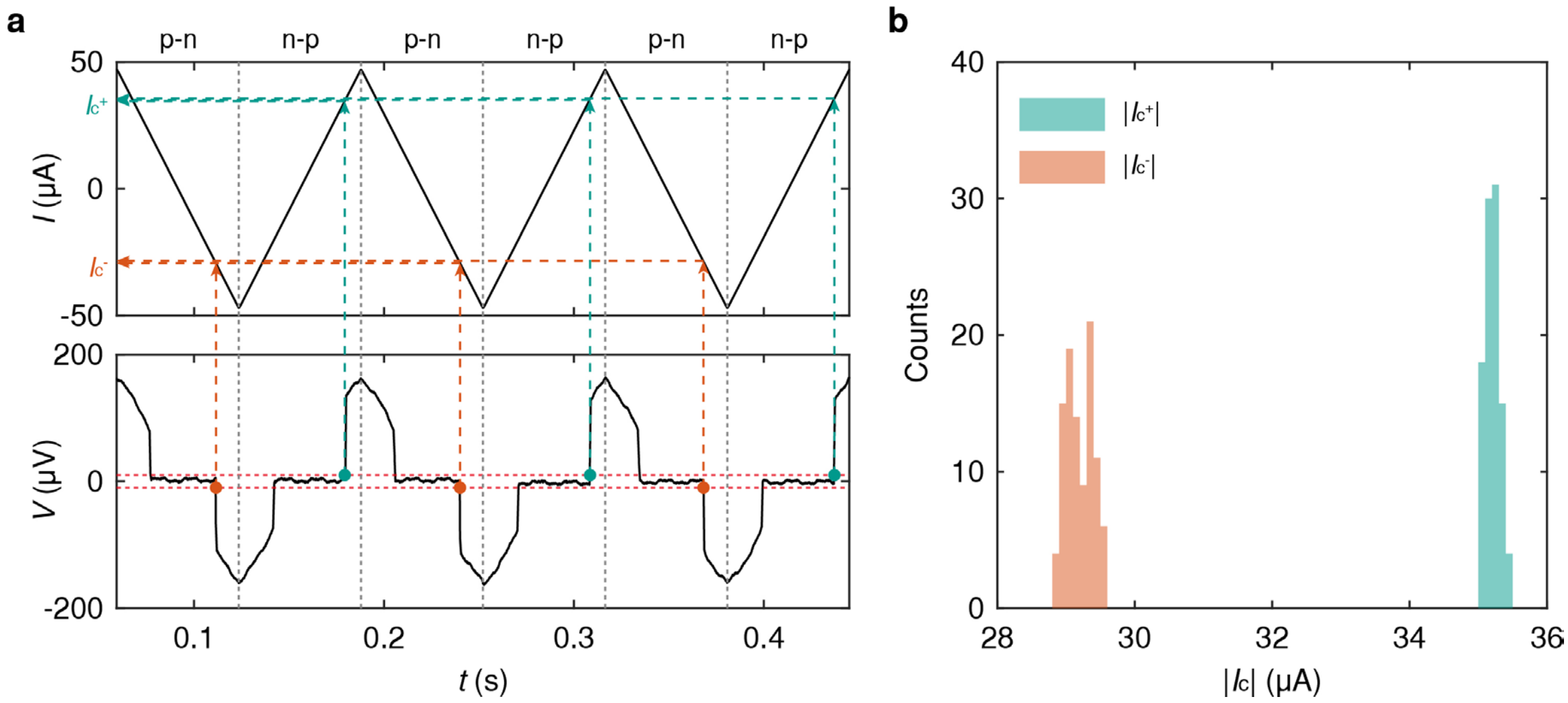

**Supplementary Fig. 2| Extraction of switching currents from triangular current bias measurements.** **a**, Applied triangular current bias $I$ (top) together with corresponding junction voltage $V$ (bottom). The p-n and n-p sweep branches are indicated at the top. Dashed red line in the lower panel indicates the voltage threshold used for switching current detection. **b**, Histograms of the extracted $|I_c^+|$ and $|I_c^-|$ values obtained from repeated current sweeps.

## Supplementary Note 3: Analysis of the Critical Current Fraunhofer Pattern

### A. Fast Oscillation Components of the Critical Current Fraunhofer Pattern

Experimental critical current Fraunhofer patterns under $x$-direction external magnetic fields exhibit an additional fast oscillation component superposed on the main Fraunhofer envelope. To separate this component and extract the background Fraunhofer pattern, we performed Fast Fourier Transform (FFT) analysis on the $\langle|I_c^+|\rangle$ and $\langle|I_c^-|\rangle$ data. As shown in Supplementary Fig. 3a, after selecting the fast oscillation range corresponding to a period of 10~15 G in the FFT spectrum, only the spectral component within that frequency window was extracted and removed from the original data. As a result, the filtered curves $\langle|I_c^+|\rangle'$ and $\langle|I_c^-|\rangle'$ were obtained for the $x$-direction.

The same procedure was applied to the $y$-direction Fraunhofer patterns. However, for the $y$-direction, the fast oscillation component isolated by FFT analysis is nearly absent, and the original $\langle|I_c^+|\rangle$ and $\langle|I_c^-|\rangle$ remain essentially identical to the filtered curves $\langle|I_c^+|\rangle'$ and $\langle|I_c^-|\rangle'$, as shown in Supplementary Fig. 3b. This suggests that the fast oscillation observed in the $x$-direction is an additional contribution distinguishable from the dominant Josephson interference across the junction. Such an additional oscillation may arise from the presence of a parallel channel capable of carrying Josephson current.

### B. Coupling with Interfacial Magnetization

In response to external magnetic fields and superconducting phase differences, an in-plane aligned magnetization $\mathbf{m}_+$ can be induced. The superconducting phase gradient, and subsequent interference effects in the current-phase relation, Eq. (S3), depends on the *total* magnetic field, which is a sum of the *external* magnetic field and the magnetic field corresponding to the in-plane aligned magnetization, i.e. $B_{x(y)} = B_{x(y)}^{\text{ext}} + \mu_0 M_s m_{+,x(y)}^{(c)}$. The aligned magnetization $\mathbf{m}_+^{(c)}$ at the critical current generically differs for positive and negative bias. Specifically, at first-order in the magnetization-supercurrent coupling strength, $m_{+,x(y)}^{(c)} = h_{x(y)}^{\text{ext}}\eta + \alpha_{x(y)}\Gamma\eta\sin\phi_c + \mathcal{O}(\Gamma^2)$, with $\phi_c$ and the corresponding $m_{+,x(y)}^{(c)}$, when substituted into Eq. (1), giving the critical current. Thus, the critical current Fraunhofer patterns as a function of the external magnetic field are laterally shifted by the corresponding in-plane aligned magnetization, giving the dominant contribution to non-reciprocal transport. The static magnetization response also effectively produces both higher harmonics in the static effective current-phase relation, and phase shifts between the harmonics, noticeable in the critical current Fraunhofer pattern. The numerical results in Supplementary Fig. 3c-d. These simulations were performed using the parameters of Eq. (S4) and

$$\frac{\omega_M}{\omega_J} \approx 0.87, \qquad \boldsymbol{\alpha} = (-15.0)\hat{\mathbf{x}} + (75.0)\hat{\mathbf{y}}, \qquad \Gamma = 4.0 \times 10^{-3}, \qquad i_2 = -0.2. \tag{S5}$$

We note the intrinsic second-harmonic amplitude $i_2$ is taken to be negative as expected for anti-aligned ferromagnets in SFFS junctions[9], similar to SNS junctions.

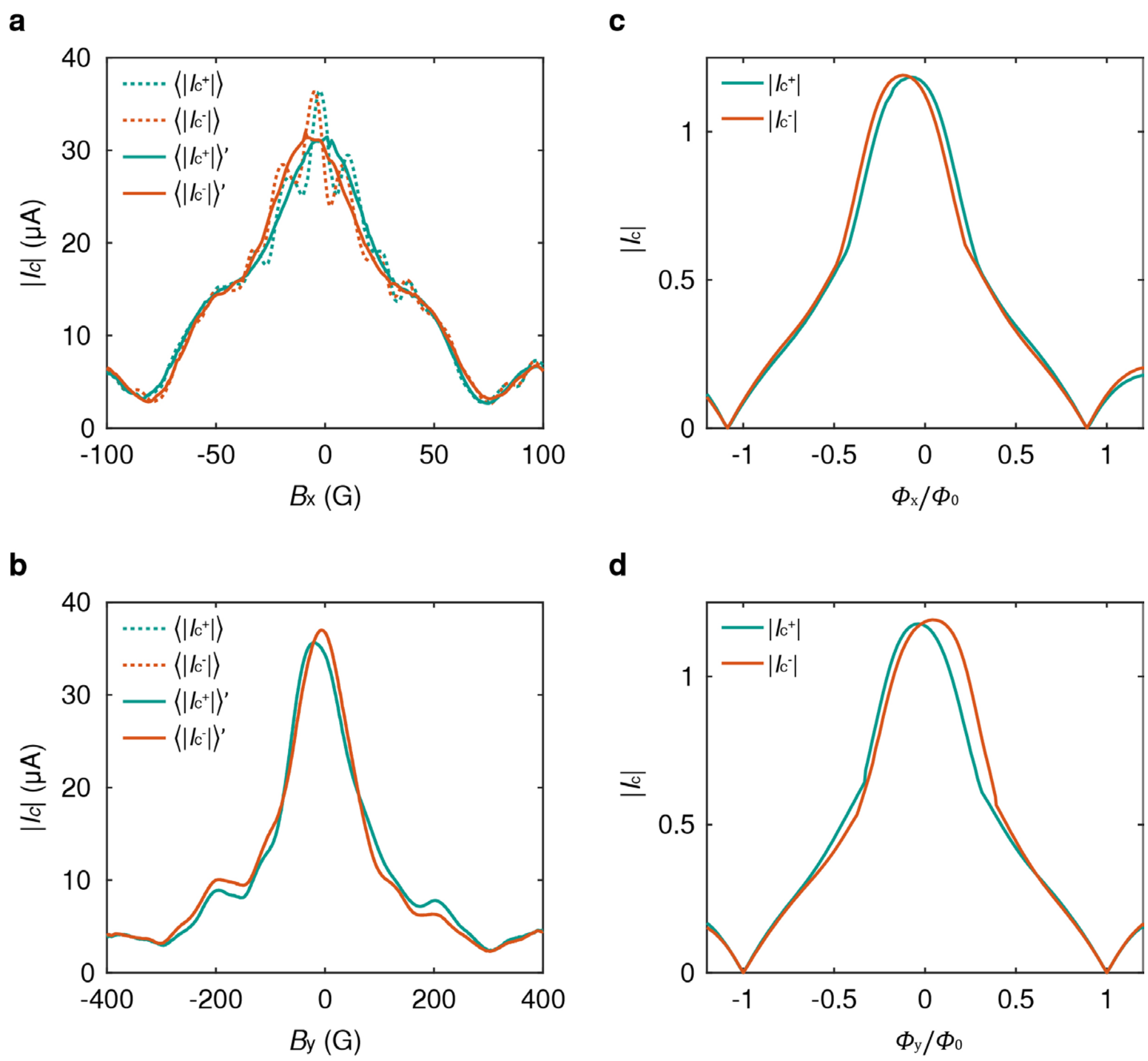

**Supplementary Fig. 3| Comparison of experimental and simulated Fraunhofer patterns. a**, **b**, Experimental Fraunhofer patterns obtained under magnetic field sweeps along the $B_{\mathrm{x}}$ and $B_{\mathrm{y}}$ directions, respectively. The dashed curves represent the original mean switching currents $\langle|I_c^+|\rangle$ (teal) and $\langle|I_c^-|\rangle$ (orange), while the solid curves denote $\langle|I_c^+|\rangle'$ and $\langle|I_c^-|\rangle'$ obtained after extracting and removing the fast oscillation component by Fast Fourier Transform (FFT) analysis. **c, d**. Simulated Fraunhofer patterns from static solutions of the coupled RCSJ + LLG equations, Eq. (S1-S3), using the parameters of Eq. (S4-S5).

## Supplementary Note 4: Analysis of the Current-Voltage Characteristics

### A. Exclusion of Fiske Resonance

Fiske resonance is a voltage step that appears when the Josephson current resonates with an internal electromagnetic cavity mode of the junction determined by its geometry[10, 11]. The resonance condition is given by

$$V_n = n\frac{h}{2e}\left(\frac{\bar{c}}{2L}\right) \tag{S6}$$

where $n$ is the cavity mode index, $\bar{c}$ is the Swihart velocity, and $L$ is the junction length perpendicular to the magnetic field direction. Accordingly, the voltage position of a Fiske step is determined by the geometric length of the junction and the electromagnetic mode velocity. Within a Fiske resonance mechanism, if the junction length $L$ changes with field orientation, the resonance voltage is also expected to change.

Figure 3 of the main text shows the hysteretic Fraunhofer pattern measured under an in-plane magnetic field sweep along $B_{\mathrm{x}}$, together with a resonance feature near $\pm V_{\mathrm{MR}}$. Correspondingly, Supplementary Fig. 4a presents the $\mathrm{d}V/\mathrm{d}I$ map and voltage bin-count map obtained from a $B_{\mathrm{y}}$ field sweep on the same junction. Under $B_{\mathrm{y}}$ sweep, a resonance feature is also observed near $\pm 170\ \mu\mathrm{V}$, which corresponds to the same energy scale as $\pm V_{\mathrm{MR}}$ identified in Fig. 3b of the main text. As shown by the optical micrograph and the schematic in Supplementary Fig. 4b, the junction length contributing to the Fraunhofer response depends on the field orientation. For $B_{\mathrm{x}}$ sweep, the effective length is approximately $17\ \mu\mathrm{m}$, whereas for $B_{\mathrm{y}}$ sweep it is about $8\ \mu\mathrm{m}$. Therefore, if the resonance were dominated by a Fiske mechanism, the resonance voltage would be expected to differ by more than a factor of two between the two field directions. In contrast, the resonance feature is observed near $\pm 170\ \mu\mathrm{V}$ for both $B_{\mathrm{x}}$ and $B_{\mathrm{y}}$ sweeps. This result suggests that the feature is difficult to explain solely in terms of a simple geometric cavity mode and is more naturally connected to the magnetic resonance (MR) induced by coupling between the interfacial magnetization and the Josephson current.

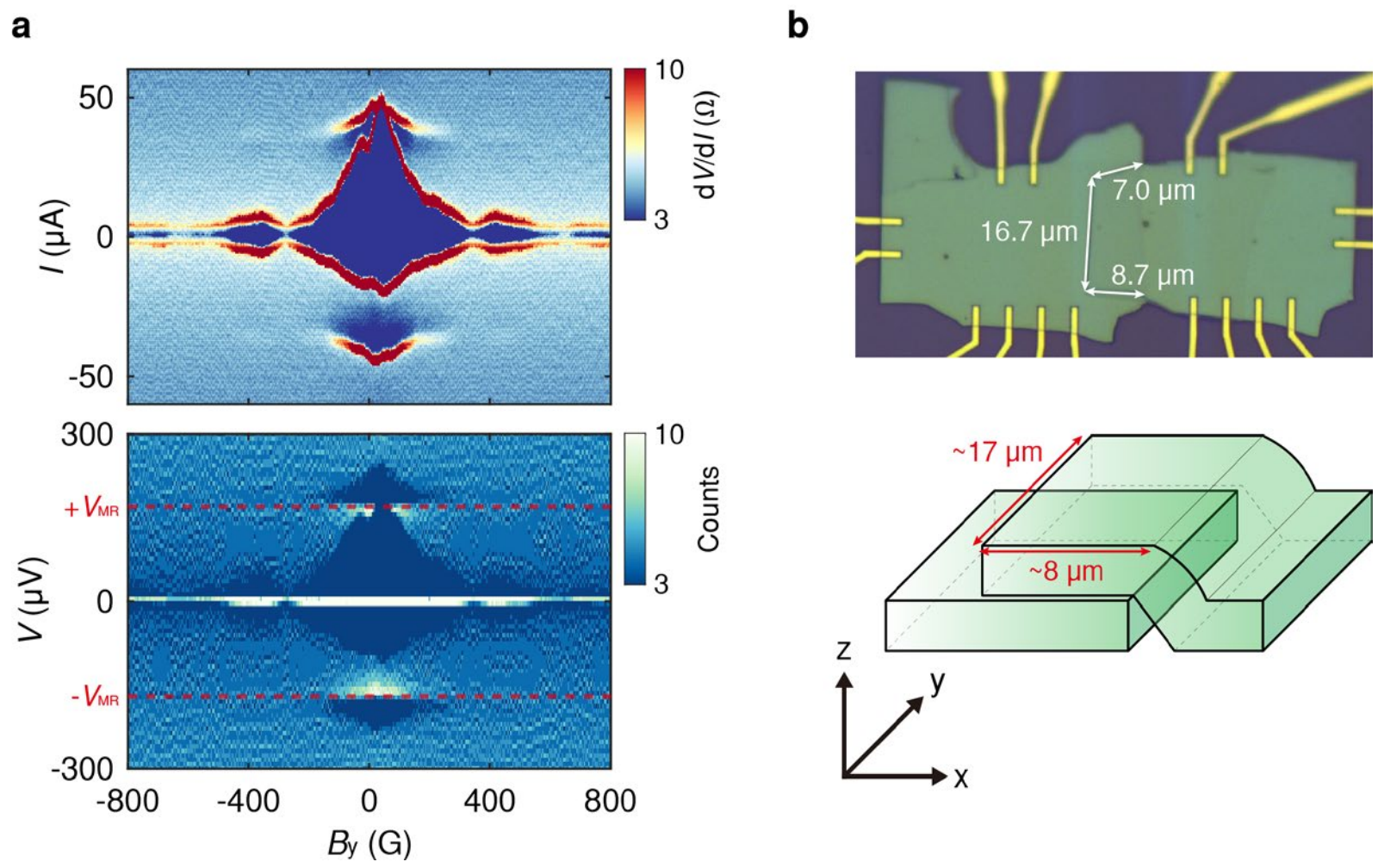

**Supplementary Fig. 4| Fraunhofer response under $B_y$ sweep and device dimensions relevant to field orientation. a**, Differential resistance $\mathrm{d}V/\mathrm{d}I$ map (top) and voltage bin-count map (bottom) of the same FTS vdW Josephson junction shown in Fig. 4, measured under an in-plane magnetic field sweep along the y direction, $B_y$. The red dashed lines mark $\pm V_{\mathrm{MR}}$, showing that a resonance feature is also observed near $\pm 170\ \mu\mathrm{V}$ under $B_y$ sweep. **b**, Optical micrograph of the device (top) and corresponding schematic (bottom). The effective junction length relevant to the Fraunhofer response depends on the field orientation, being approximately $17\ \mu\mathrm{m}$ for $B_x$ sweep and about $8\ \mu\mathrm{m}$ for $B_y$ sweep.

## B. Magnetic Resonance (MR)

Here we present numerical simulations combining the LLG and RCSJ equations, which well reproduce MR in the current–voltage characteristics. Numerical simulations of the full RCSJ + LLG equations were performed using the parameters of Eq. (S4) and

$$\frac{\omega_M}{\omega_J} \approx 0.87, \qquad \boldsymbol{\alpha} = (-6.0)\hat{\mathbf{x}} + (30.0)\hat{\mathbf{y}}, \qquad \Gamma = 4.0 \times 10^{-3}, \qquad \beta_J = 7.5, \qquad i_2 = -0.1. \qquad \text{(S7)}$$

We note $\boldsymbol{\alpha}$ is suppressed relative to Eq. (S5) by dynamic screening.

For weak Josephson current coupling, the anomalous conductance appears at $V_{\mathrm{DC}} \approx \pm V_{\mathrm{MR}} = \pm(2e/\hbar)^{-1}\omega_M$, at a bandwidth which scales with $\sim \zeta_M \eta^{-1} V_{\mathrm{MR}}$. Depending on the parameters, voltage jumps can appear with hysteresis between positive and negative current biases, explaining the voltage jump, and the smooth crossover from voltage jumps to merely deviations in differential conductance (Fig. 3a). As the magnetic characteristic frequency (natural frequency), and the corresponding characteristic voltage is set by material properties, namely the interlayer antiferromagnetic exchange coupling between the top and bottom layer and the magnetic anisotropies, it does not exhibit dependence on external magnetic field directions, consistent with our experiments. Additionally, the anomalous conductance is strongly affected by interference effects, exhibiting a proportionality to $\left(\frac{\sin(\pi\Phi_x/\Phi_0)}{\pi\Phi_x/\Phi_0}\right)^2$ at lowest order in the underdamped limit ($\beta_J \to \infty$)[7], unlike Fiske resonance. The presence of strong interference effects is also consistent with our experiments.

## Supplementary Note 5: Additional Shapiro Step Measurements

Weak dissipation in superconducting phase dynamics generally leads to the emergence of fractional Shapiro steps beyond what is reflective of the effective current-phase relation, even irrational Shapiro steps and chaotic dynamics, as well as overlapping Shapiro steps and hysteresis[12]. In particular, inertial effects ($\beta_J > 0$) naturally give rise to half-integer Shapiro steps, although they only become pronounced for underdamped dynamics. It has also been reported that when $\omega_M \approx \omega_J \approx \Omega_S$, Shapiro steps may be strongly renormalized and chaotic[13, 14]. In Supplementary Fig. 5, we present the microwave response at $B_x = 0$ G, which exhibits many fractional Shapiro steps, some of which appear to arise directly from MR, and exhibit strong hysteresis, making them undesirable for analyzing the effective current-phase relation.

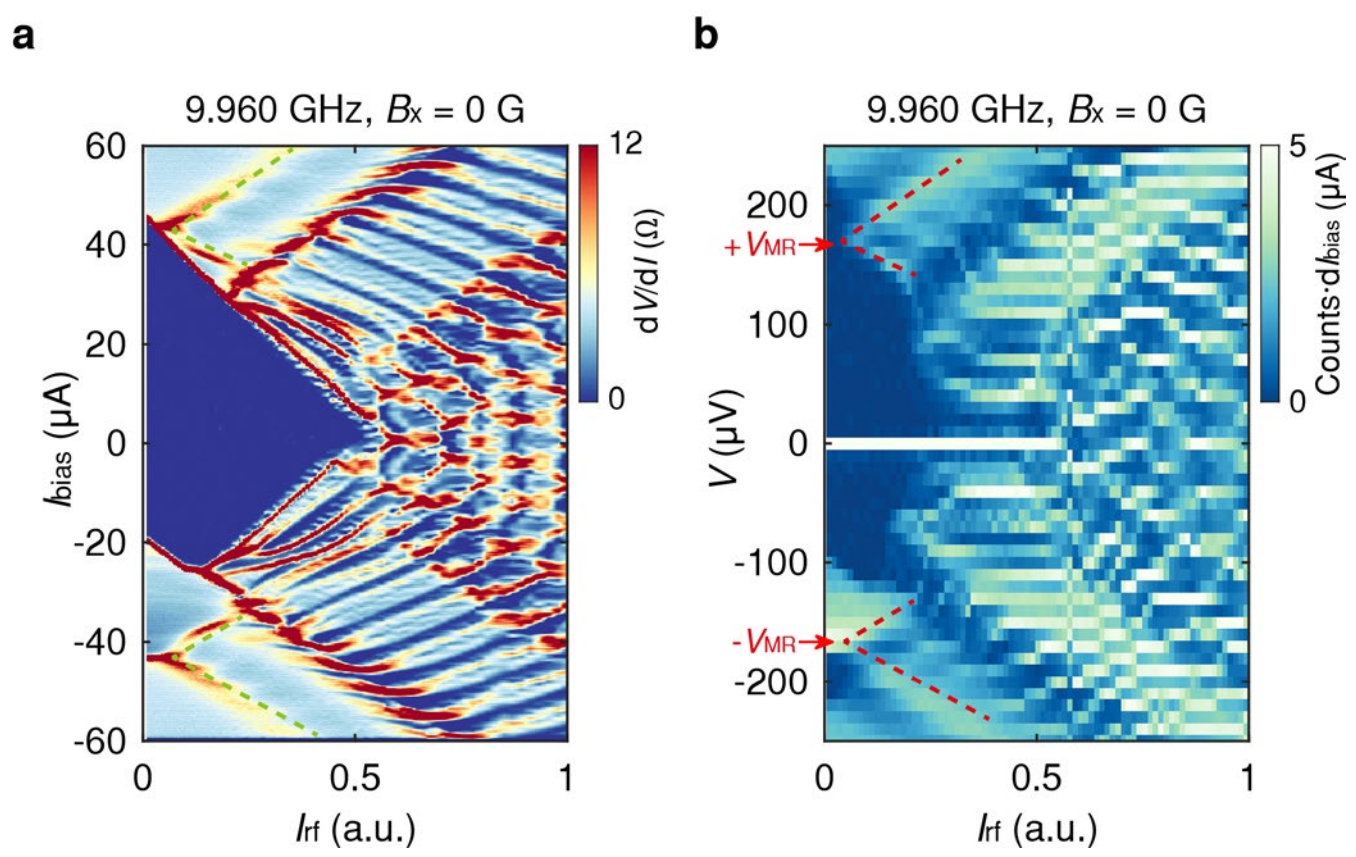

**Supplementary Fig. 5| Microwave response at $B_x = 0$ G. a**, Differential resistance $\mathrm{d}V/\mathrm{d}I$ as a function of dc bias current $I_{\mathrm{bias}}$ and RF drive amplitude $I_{\mathrm{rf}}$ measured at 9.960 GHz and $B_x = 0$ G. The green dashed lines guide additional step features arising from MR. **b**, Bin-count map constructed from the data in (**a**), plotted as a function of voltage $V$ and RF drive amplitude $I_{\mathrm{rf}}$. The red arrows indicate the characteristic voltages $\pm V_{\mathrm{MR}}$, and the red dashed lines guide the unintended additional step features formed by the overlap of MR and the external RF response.

We perform Shapiro step analysis near the node of the Fraunhofer pattern, $B_x = 75\ \mathrm{G}$. As shown in Supplementary Fig. 6a-d, under this condition the difference between the n-p and p-n sweeps is negligible and the $I - V$ hysteresis is essentially absent. Supplementary Fig. 6 shows additional Shapiro step data measured at $B_\mathrm{x} = 75\ \mathrm{G}$. Both the $\mathrm{d}V/\mathrm{d}I$ maps and the bin-count maps measured at 4.348, 7.580, and 9.680 GHz consistently exhibit integer and asymmetric half-integer Shapiro step structures.

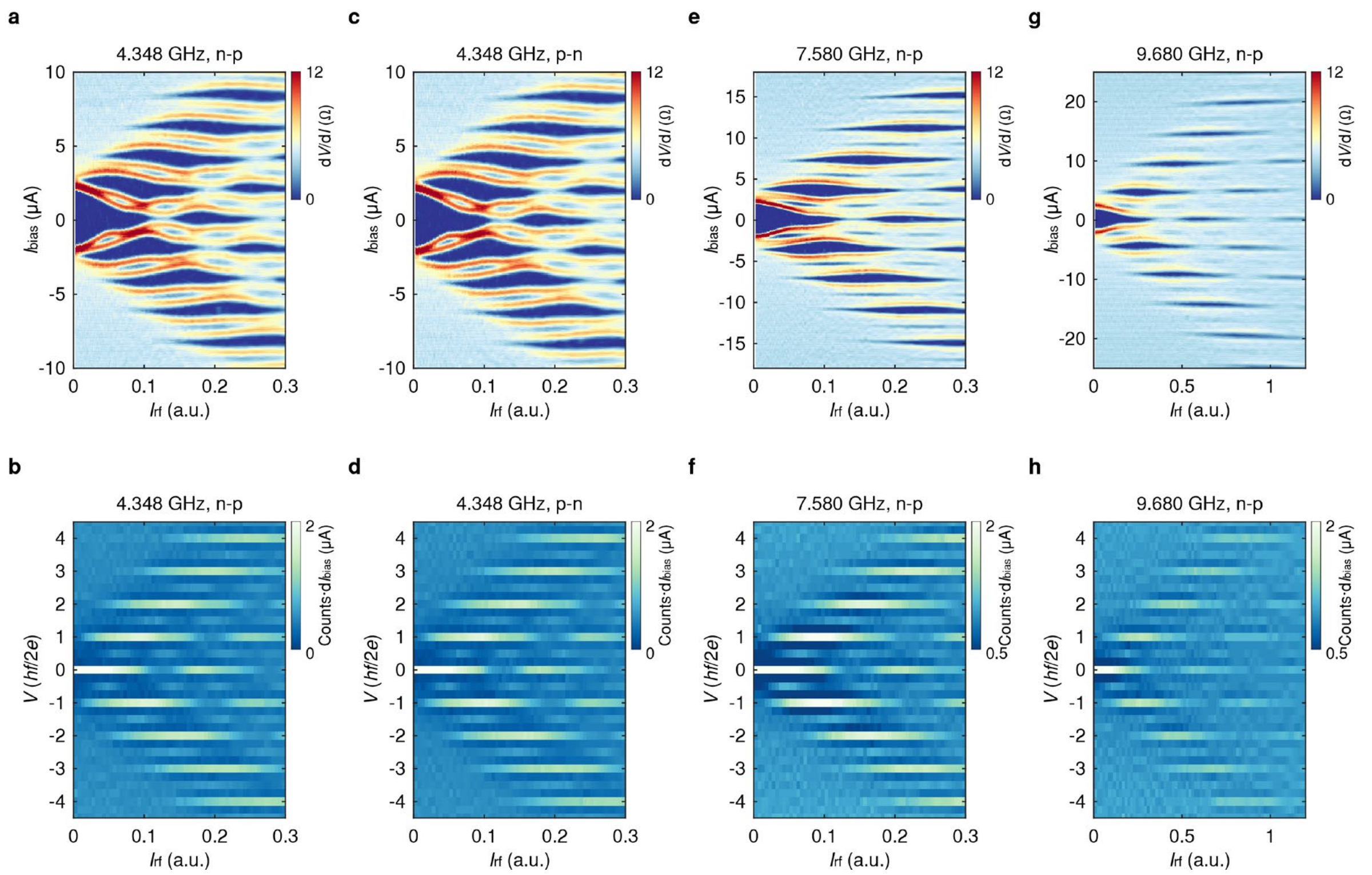

**Supplementary Fig. 6| Additional Shapiro steps measurements at $B_\mathrm{x} = 75\ \mathrm{G}$ under different microwave frequencies. a**, Differential resistance $\mathrm{d}V/\mathrm{d}I$ as a function of dc bias current $I_\mathrm{bias}$ and RF drive amplitude $I_\mathrm{rf}$ measured at 4.348 GHz during the n-p current sweep. **b**, Bin-count map constructed from the data in (**a**), plotted as a function of voltage $V$ normalized by $hf/2e$ and RF drive amplitude $I_\mathrm{rf}$. **c**, **d**, Corresponding $\mathrm{d}V/\mathrm{d}I$ map and bin-count map measured at the same frequency during the p-n current sweep. **e-h**, Additional $\mathrm{d}V/\mathrm{d}I$ maps and corresponding bin-count maps measured at 7.580 GHz and 9.680 GHz.

Supplementary Fig. 7 provides additional analysis of the 5.776 GHz data shown in Fig. 4 of the main text. The same dataset is replotted using the microwave power scale, and the step evolution is directly examined through line cuts taken at selected power values. Supplementary Fig. 7b shows the clear formation of both integer and half-integer Shapiro steps. In addition, extra line cuts of the bin-count map are presented for integer and half-integer steps, showing that the asymmetric half-integer Shapiro response emphasized in the main text are not limited to a single step, but appear consistently in other half-integer steps as well.

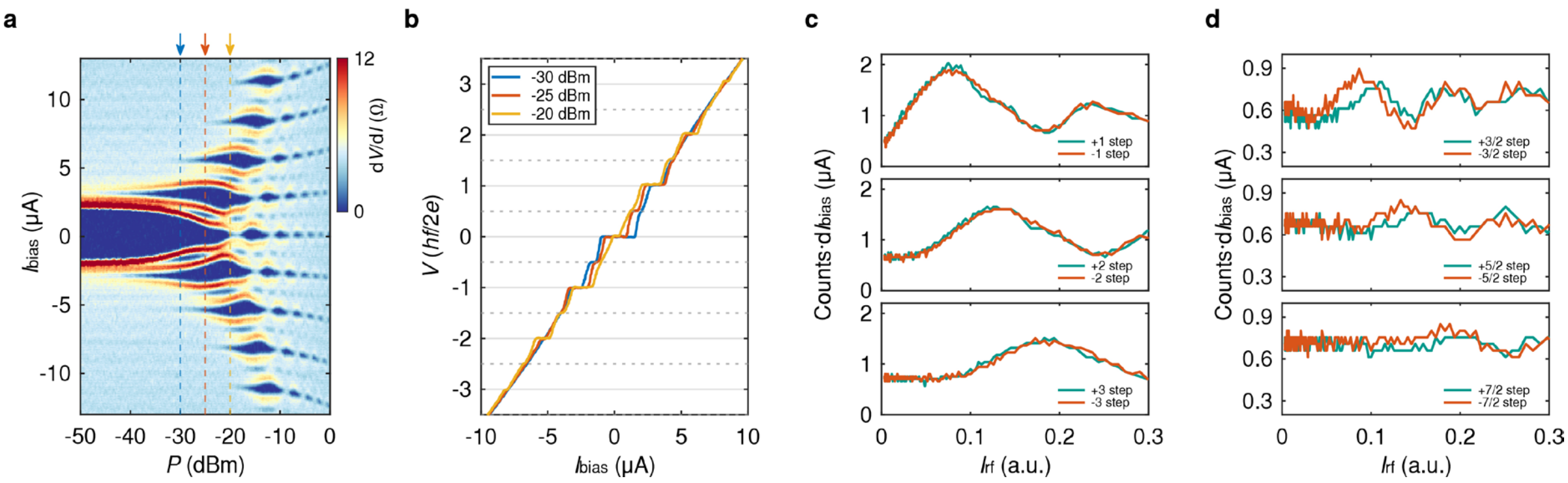

**Supplementary Fig. 7| Additional analysis of the Shapiro steps at 5.776 GHz and $B_x = 75$ G. a**, Differential resistance $dV/dI$ as a function of dc bias current $I_{bias}$ and RF power $P$, replotted from the same dataset as Fig. 3c using the raw power scale. Dashed vertical lines mark the power values used for the line cuts in (**b**). **b**, Line cuts of the data in (**a**) taken at $P = -30$, $-25$, and $-20$ dBm, showing the voltage response $V$ normalized by $hf/2e$ as a function of $I_{bias}$. **c**, Line cuts of the bin-count map in Fig. 3d taken at the integer Shapiro steps $n = \pm1, \pm2$, and $\pm3$ as a function of $I_{rf}$. **d**, Line cuts of the bin-count map in Fig. 3d taken at the half-integer Shapiro steps $n = \pm3/2, \pm5/2$, and $\pm7/2$ as a function of $I_{rf}$.

## Supplementary Note 6: Analysis of the Shapiro Steps

To analyze the Shapiro-step response, we first employ the RCSJ model, Eq. (S2), without explicitly including the magnetic degree of freedom. Instead, we consider an effective CPR consisting of a fundamental Josephson term, a second-harmonic term, and a phase-shift term as

$$I^{\mathrm{eff}}(\phi) = I_1 \sin\left(\phi - \phi_0^{\mathrm{eff}}\right) + I_2^{\mathrm{eff}} \sin 2\phi \,, \tag{S8}$$

where $I_2^{\mathrm{eff}}$ is the effective second-harmonic amplitude, and $\phi_0^{\mathrm{eff}}$ is the effective phase shift.

As confirmed by the absence of current-bias hysteresis in the Shapiro measurements at $B_{\mathrm{x}} = 75\ \mathrm{G}$ (Supplementary Fig. 6), the experimental data were taken in the overdamped regime. Accordingly, the Shapiro steps were simulated with the overdamped RCSJ equations,

$$\frac{\mathrm{d}\phi}{\mathrm{d}\left(\omega_J t\right)} = -i^{\mathrm{eff}}(\phi) + i_{\mathrm{dc}} + i_{\mathrm{ac}} \sin(\omega_{\mathrm{RF}} t)\,. \tag{S9}$$

Here $i^{\mathrm{eff}}(\phi) = I^{\mathrm{eff}}(\phi)/I_1$ is the normalized effective CPR, Eq. (S8), $i_{\mathrm{dc}} = I_{\mathrm{bias}}/I_1$, $i_{\mathrm{ac}} = I_{\mathrm{rf}}/I_1$ are the DC and AC current amplitudes, and $\omega_{RF}$ is the RF drive angular frequency.

To establish the role played by the coupling with the interfacial magnetization, additional control simulations are presented in Supplementary Fig. 8. For $i_2^{\mathrm{eff}} = 0$, only integer Shapiro steps appear, with no half-integer steps. While half-integer Shapiro steps of comparable size to experiments do emerge for $i_2^{\mathrm{eff}} = 0.7$, this is an unrealistically large amplitude for predominantly $s$-wave superconductivity, and does not produce asymmetric half-integer Shapiro steps. Therefore, the control simulations show that a second-harmonic component alone is insufficient. Additionally, an effective phase-shift component, $\phi_0^{\mathrm{eff}} \approx -0.29$, is required to reproduce the experimentally observed asymmetric half-integer response (Fig. 4d-f).

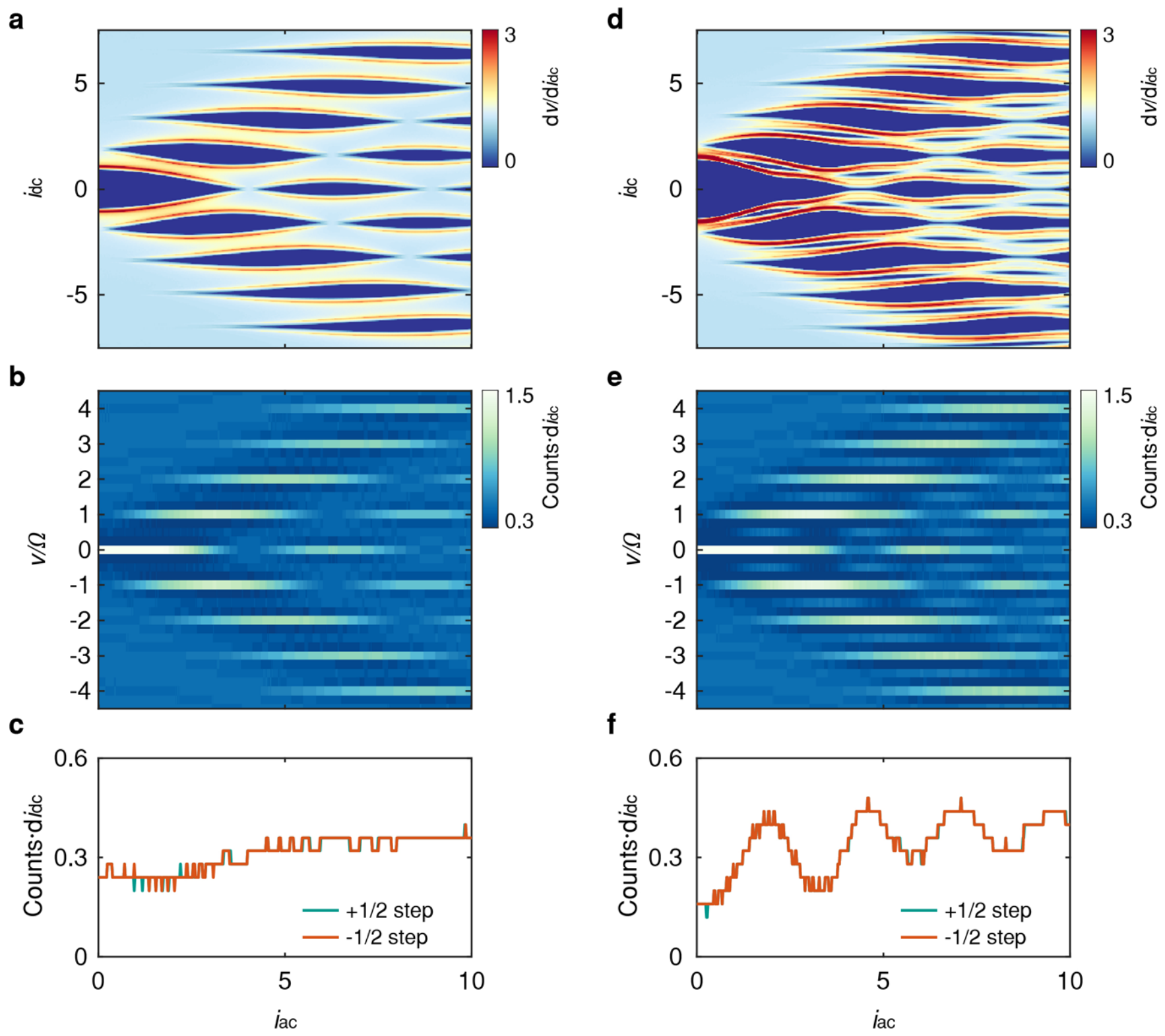

**Supplementary Fig. 8| Shapiro steps without coupling with interfacial magnetization.** Simulated $\mathrm{d}V/\mathrm{d}I$ map, bin-count map, and $\pm 1/2$ step line cuts for **a-c**, $i_2^{\mathrm{eff}} = 0$, **d-f**, $i_2^{\mathrm{eff}} = 0.7$.

This phase-shift component can arise either from magnetic-field-induced interfacial magnetization or from the time lag of the magnetization relative to the superconducting phase. The former is dominant for our experimentally studied regime of $\omega_M \gg \omega_{RF}$ for lower Shapiro steps[7], while the dynamic response is primarily relevant for $\omega_M \approx \omega_J \approx \omega_{RF}$ [13, 14]. This comparison shows that coupling with interfacial magnetization is necessary to reproduce key features of the experimental microwave response.

Numerical simulations of the full LLG + overdamped RCSJ equations (Fig. 4g-i) were performed using the parameters of Eq. (S4) and

$$\frac{\omega_{RF}}{\omega_J} \approx 1.6, \qquad \frac{\omega_M}{\omega_J} \approx 22.6, \qquad \boldsymbol{\alpha} = (40.0)\hat{\mathbf{x}} + (20.0)\hat{\mathbf{y}}, \qquad \begin{cases} \Gamma_x = 2.0 \times 10^{-3} \\ \Gamma_y = 0.2 \times 10^{-3} \end{cases}, \qquad i_2 = 0.1, \qquad \text{(S10)}$$

taking into account anisotropic interference-rescaling of parameters due to being proximate to the critical current Fraunhofer node, i.e. $\Phi_x \approx \Phi_0$[7].

**References**

1. Shan W-Y, Lu H-Z, Shen S-Q. Effective continuous model for surface states and thin films of three-dimensional topological insulators. *New Journal of Physics* 2010, **12**(4)**:** 043048.
2. Qiu G, Yang H-Y, Hu L, Zhang H, Chen C-Y, Lyu Y*, et al.* Emergent ferromagnetism with superconductivity in Fe(Te,Se) van der Waals Josephson junctions. *Nat Commun* 2023, **14**(1)**:** 6691.
3. Bernevig BA, Zhang S-C. Spin splitting and spin current in strained bulk semiconductors. *Phys Rev B* 2005, **72**(11)**:** 115204.
4. Li F, Zhang Q, Tang C, Liu C, Shi J, Nie C*, et al.* Atomically resolved FeSe/SrTiO3(001) interface structure by scanning transmission electron microscopy. *2d Mater* 2016, **3**(2)**:** 024002.
5. Fukaya Y, Zhou G, Zheng F, Zhang P, Wang L, Xue Q-K*, et al.* Asymmetrically optimized structure in a high-Tc single unit-cell FeSe superconductor. *Journal of Physics: Condensed Matter* 2019, **31**(5)**:** 055701.
6. Buzdin A. Direct Coupling Between Magnetism and Superconducting Current in the Josephson $\varphi_0$ Junction. *Phys Rev Lett* 2008, **101**(10)**:** 107005.
7. Ahn C-E, Shin S-H, Lee G-H, Cho GY. in preparation.
8. Konschelle F, Buzdin A. Magnetic Moment Manipulation by a Josephson Current. *Phys Rev Lett* 2009, **102**(1)**:** 017001.
9. Pajović Z, Božović M, Radović Z, Cayssol J, Buzdin A. Josephson coupling through ferromagnetic heterojunctions with noncollinear magnetizations. *Phys Rev B* 2006, **74**(18)**:** 184509.
10. Coon DD, Fiske MD. Josephson ac and Step Structure in the Supercurrent Tunneling Characteristic. *Physical Review* 1965, **138**(3A)**:** A744-A746.
11. Fiske MD. Temperature and Magnetic Field Dependences of the Josephson Tunneling Current. *Rev Mod Phys* 1964, **36**(1)**:** 221-222.
12. Kautz RL. Noise, chaos, and the Josephson voltage standard. *Rep Prog Phys* 1996, **59**(8)**:** 935.
13. Zhai B, Li B, Wen Y, Wu F, He J. Prediction of ferroelectric superconductors with reversible superconducting diode effect. *Phys Rev B* 2022, **106**(14)**:** L140505.
14. Botha AE, Shukrinov YM, Tekić J, Kolahchi MR. Chaotic dynamics from coupled magnetic monodomain and Josephson current. *Physical Review E* 2023, **107**(2)**:** 024205.